\documentclass[12pt]{emulateapj}
\usepackage{epsf}
\usepackage{lscape}

\newcommand{\rmg}{${\cal R}_{\rm mg}$}
\newcommand{\tmg}{$T_{\rm mg}$}
\newcommand{\cmg}{$C_{\rm mg}$}
\newcommand{\fpair}{$f_{\rm pair}$}

\newcommand{\kband}{$K_s$-band}
\newcommand{\K}{$K_s$}
\newcommand{\photoz}{photo-$z$}
\newcommand{\photozs}{photo-$z$s}

\newcommand{\msun}{M$_{\odot}$}

\shorttitle{The Greater Impact of Mergers on the Growth of
   Massive Galaxies}
\shortauthors{Bundy, Fukugita, Ellis, Targett, Belli, Kodama}

\submitted{Submitted to ApJ and revised after referee comments}

\begin{document}

\title{The Greater Impact of Mergers on the Growth of Massive Galaxies: Implications for
  Mass Assembly and Evolution Since $z \simeq 1$\footnotemark[*]}

\footnotetext[*]{Based on observations collected at the Subaru
Telescope, which is operated by the National Astronomical Observatory of
Japan, and with the NASA/ESA HST, obtained at STScI, which is operated
by AURA, under NASA contract NAS5-26555.}

\author{Kevin Bundy\altaffilmark{1,2}, Masataka Fukugita\altaffilmark{3}, Richard S. Ellis\altaffilmark{4,5}, Thomas A. Targett\altaffilmark{4}, 
Sirio Belli\altaffilmark{6}, Tadayuki Kodama\altaffilmark{7}}

\altaffiltext{1}{Reinhardt Fellow, Department of Astronomy, University of Toronto, 
60 George Street, Toronto ON M5S 3H8, Canada}
\altaffiltext{2}{Hubble Fellow, Astronomy Department, University of
  California, Berkeley, CA 94705}
\altaffiltext{3}{Institute for Cosmic Ray Research, University of Tokyo,
Kashiwa 2778582, Japan}
\altaffiltext{4}{Astronomy, Mail Stop 105-24, California Institute
of Technology, Pasadena CA 91125}
\altaffiltext{5}{Department of Astrophysics, Keble Road, Oxford OX1 3RH,
UK}
\altaffiltext{6}{Universit\'a degli Studi di Bologna, via Ranzani 1, Bologna
I-40127, Italy}
\altaffiltext{7}{National Astronomical Observatory of Japan, Mitaka,
Tokyo 181--8588, Japan}

\begin{abstract}

  {Using deep infrared observations conducted with the MOIRCS imager on
    the Subaru Telescope in the northern GOODS field combined with
    public surveys in GOODS-S, we investigate the dependence on stellar
    mass, $M_*$, and galaxy type of the close pair fraction ($5 h^{-1}
    {\rm ~kpc} < r_{\rm sep} < 20 h^{-1} {\rm ~kpc}$) and implied merger
    rate.  In terms of combined depth and survey area, our publicly
    available mass-limited sample represents a significant improvement
    over earlier infrared surveys used for this purpose.  In common with
    some recent studies we find that the fraction of paired systems that
    could result in major mergers is low ($\sim$4\%) and does not
    increase significantly with redshift to $z \approx 1.2$, with
    $\propto (1+z)^{1.6 \pm 1.6}$.  Our key finding is that massive
    galaxies with $M_* > 10^{11}$\msun\ are more likely to host merging
    companions than less massive systems ($M_* \sim 10^{10}$\msun).  We
    find evidence for a higher pair fraction for red, spheroidal hosts
    compared to blue, late-type systems, in line with expectations based
    on clustering at small scales.  So-called ``dry'' mergers between
    early-type galaxies devoid of star formation represent nearly 50\%
    of close pairs with $M_* > 3 \times 10^{10}$\msun\ at $z \sim 0.5$,
    but less than 30\% at $z \sim 1$.  This result can be explained by
    the increasing abundance of red, early-type galaxies at these
    masses.  We compare the volumetric merger rate of galaxies with
    different masses to mass-dependent trends in galaxy evolution.  Our
    results reaffirm the conclusion of \citet{bundy07} that major
    mergers do not fully account for the formation of spheroidal
    galaxies since $z \sim 1$.  In terms of mass assembly, major mergers
    contribute little to galaxy growth below $M_* \sim 3 \times
    10^{10}$\msun\ but play a more significant role among galaxies with
    $M_* \gtrsim 10^{11}$\msun\, $\sim$30\% of which have undergone
    mostly dry mergers over the observed redshift range.  Overall, the
    relatively rapid and recent coalescence of high mass galaxies
    mirrors the expected hierarchical growth of halos and is consistent
    with recent model predictions, even if the top-down suppression of
    star formation and morphological evolution (i.e., ``downsizing'')
    involves additional physical processes.}

\end{abstract}

\keywords{galaxies: evolution --- galaxies: formation --- galaxies: interactions}

\section{Introduction}\label{intro}

Recent surveys have challenged the traditional view that merging between
dark matter halos governs the mass assembly history and
evolution of the galaxies hosted by these halos.  The surprising abundance of
established massive galaxies at $z \simeq 2$ \citep[e.g.,][]{glazebrook04,
  cimatti04} has been followed by increasing evidence for an early
completion of star formation in the most massive galaxies, followed by
continued activity in lower-mass systems---a phenomenon now termed
``downsizing'' \citep[e.g.,][]{juneau05, treu05, bundy06, borch06,
  cimatti06, bell07, cowie08}.  These observations reveal top-down
evolutionary patterns that stand in contrast to the expected
hierarchical, or bottom-up, nature of CDM halo assembly.

While semi-analytic models employing various energy feedback
prescriptions have moved closer to reproducing top-down behavior within
the hierarchical framework \citep{croton06, bower06, de-lucia06,
  cattaneo08, stringer08}, it has not been observationally possible to
verify the CDM prediction that via merging, more massive galaxies are
assembled at later times.  In principle, the signal is encoded in the
evolving stellar mass function (MF) which would be expected to show a
rising number density at the high-mass end as a function of cosmic time.
Current surveys, however, are too much affected by cosmic variance (or,
more accurately, sample variance) uncertainties to confirm this trend
\citep[see][]{stringer08}.  It is also difficult to separate evolution
in the MF resulting from galaxy mergers from that associated with star
formation (SF) \citep{drory08}.  The clear alternative is to study
galaxy mergers directly and to quantify their role in driving mass
assembly by determining how the galaxy merger rate depends on mass.

In addition to driving mass assembly, major mergers (defined here to be
those involving components with a mass
ratio greater than $1/4$) have also been proposed as the key mechanism
that transforms disk-like galaxies into spheroidals \citep{toomre77}.
It may also shut off star formation via triggered AGN feedback
\citep[see][]{hopkins07a}.  Mergers may not only move galaxies onto the
``red sequence,'' but so-called ``dry mergers'' between red-sequence
systems have been invoked to explain the increasing contribution at $M_*
> 10^{10}$\msun\ to the global stellar mass density from red galaxies.
\citep[e.g.,][]{van-dokkum05a, faber07, bell07}.  

The fundamental question is whether the {\em major} merger rate is high
enough to explain the mass-dependent increase in the numbers of
spheroidal and red-sequence galaxies or whether other mechanisms for
building spheroidal systems are needed.  \citet{bundy07} appealed to
numerical simulations to argue that the predicted major merger rate
among dark matter halos is too low.  A similar conclusion was also
reached by \citet{genel08}.  It is now imperative to compare the
formation rate of spheroidals to observations capable of identifying
major mergers and resolving their frequency as a function of mass.

Previous observational attempts to measure the galaxy merger rate since
$z \sim 1$ have faced a number of challenges and sometimes disagree in
their conclusions.  The derived rate depends on the assumed efficiency
and merger timescale, but more importantly, merger definitions vary
among authors.  Additionally, it has been difficult to distinguish major
from minor mergers.  At $z \lesssim 1$, the merger rate has been
estimated based on the occurrence of morphologically disturbed systems
\citep[e.g.,][]{conselice03a, lotz08} and the frequency of optical
pairs, either defined with respect to their relative velocity difference
\citep[e.g.,][]{patton02, lin04, lin08, de-ravel08} or corrected for chance
projections \citep[e.g.,][]{le-fevre00, bell06a, kartaltepe07}.
Parameterizing the redshift dependence as $\propto (1+z)^m$, the range of
reported values is virtually unconstrained, with $m = 0$--4.  Part of
the problem may be the use of optical samples which we argued in
previous work based on a near-IR sample \citep{bundy04}, may be biased
by triggered SF \citep[also see][]{berrier06}.

Although great care is needed in making comparisons with these methods,
each of which defines a ``merger'' in a different way, it is now clear
that there is considerable uncertainty in the literature regarding both
the rate of merging and whether it rises substantially with redshift.
Our goal in this paper is to overcome some of the problems faced in
previous work by using a mass-limited sample of galaxy pairs drawn from
the Great Observatories Origins Deep Survey fields
\citep[GOODS,][]{giavalisco04}.  Our \K-selected sample is more than 15
times larger than that discussed in \citet{bundy04}.  The enlarged
sample allows us to identify pairs using the physically motivated
definition of a major merger---as determined by the inferred mass
ratio---and therefore enables us to measure the impact of mergers on
mass assembly and make direct comparisons to evolutionary trends such as
the formation rate of spheroidal galaxies.  With our large sample size,
we can study the merger rate as a function of mass, a key tool for
testing the late assembly times predicted for massive galaxies.  We can
also distinguish dry and ``wet'' mergers (those involving significant
cold gas reservoirs) as well as the types of galaxies most likely to
host close companions.  This project combines publicly available data in
GOODS-S with new observations obtained in GOODS-N with the
recently-commissioned Multi-Object Infrared Camera and Spectrograph
\citep[MOIRCS,][]{ichikawa06} panoramic infrared imager on the Subaru
8.2m telescope.

Using this data set, we address two important questions in this paper.
The first is whether the rate of major mergers since $z \sim 1$ is
sufficient to explain the formation of new spheroidal and red-sequence
galaxies over this redshift range.  The second goal is to quantify the
role of mergers in galaxy growth and determine whether the mass
dependence of the merger rate is consistent with expectations based on
hierarchical assembly.

A plan of the paper follows. In \S\ \ref{data} we introduce the new
MOIRCS dataset and discuss its image processing and simulations undertaken
to determine its limiting magnitude.  We also present our analysis of
complementary near-IR data that is publicly available and was obtained
in GOODS-S using the Infrared Spectrometer and Array Camera (ISAAC) on
the Very Large Telescope (VLT).  The resulting \kband\ catalogs are correlated with a number of
public redshift and imaging surveys, and the matched catalogs will be
made available online.  In \S\ \ref{properties}, we
discuss our estimates of stellar masses, rest-frame colors, and
morphologies.  We present our methods for measuring pair fractions and
comoving number densities in \S\ \ref{methods}.  The results are
described in \S\ \ref{results}, while in \S\ \ref{discussion} we discuss the
inferred merger rates and implications for both mass assembly and
morphological/color transformations.  A summary is provided in \S\ \ref{summary}.

Throughout this paper, we use the AB magnitude system and adopt a
standard cosmology with $H_0$=70 $h_{70}$ km s$^{-1}$ Mpc$^{-1}$,
$\Omega_M$=0.3 and $\Omega_{\Lambda}$=0.7. 

\section{Data and Catalogs}\label{data}

Our study of galaxy pairs makes use of the GOODS fields, which not only
provide deep HST imaging in four bands, but have also been targeted by
many followup surveys at a variety of wavelengths.  We begin this
section with a description of the acquisition and reduction of our new
MOIRCS data that delivers much-needed deep \kband\ coverage across the
entire GOODS-N.  We then turn to publicly available data, presenting
first our analysis of near-IR imaging in GOODS-S obtained with ISAAC.
These data are very similar in depth and resolution to the MOIRCS data,
and together the two datasets provide the \K-detected source lists that
form the basis of the samples used in this paper.

We then discuss how we cross reference our \K-selected catalogs to the
many imaging and spectroscopic datasets that are available in both GOODS
fields.  Because a key part of our merger rate analysis uses redshift
information to confirm potential galaxy pairs, it is important to make
use of as many spectroscopic redshifts as possible.  For this reason, we
match our sample to the most up-to-date versions of available surveys
including many of the recently completed ESO surveys in
GOODS-S\footnote{In GOODS-S, our catalog serves as a partial update to
  the MUSIC compilation of \citet{grazian06}.}.  Finally, where
previously measured spectroscopic or photometric redshifts are not
available, we describe the additional photometric redshifts that are
required to supplement our sample.  Recognizing the value of \K-selected
multiwavelength compilations in GOODS, the final matched catalogs will
become publicly available at this website in early 2009: {\tt
 astro.berkeley.edu/$\sim$kbundy/KGOODS/}

\subsection{MOIRCS Near-IR Imaging}

We obtained \kband\ observations using  MOIRCS on the Subaru
Telescope during two nights in April 2006.  Each MOIRCS pointing is
imaged onto two overlapping detectors, giving a total field of view that
is roughly 4\arcmin\ by 7\arcmin.  Our observations consisted of 8 tiled
pointings arranged with some overlap in order to cover the entire
GOODS-N region.  At each pointing, we executed three to four 9-point
dither patterns (dithering by 15\arcsec), coadding 4 exposures at each
position.  Individual exposure times varied from 50 to 55 seconds.  The
total integration time was typically one hour across the field with average
seeing of 0\farcs5.  The MOIRCS \kband\ 80\% completeness depth varies
across the GOODS-N field from $K_s=23.5$ to $K_s=24.0$ with an average
of $K_s \approx 23.8$. 

We reduced the data using the MCSRED IRAF package written by Ichi
Tanaka.  Because MOIRCS is a relatively new instrument, we report in the
Appendix on our modifications to MCSRED and solutions to other problems
we encountered during the data reduction.  Once the final images were
reduced and combined we registered them to the GOODS-N ACS astrometry by comparing
bright stars detected in both the MOIRCS and ACS data.  Our Subaru
observations were not taken in completely photometric conditions, and so
photometric calibration was carried out by comparing to stars in the
shallower ($K_s \lesssim 22.7$) Palomar \kband\ observations
presented in \citet{bundy05}.

We used SExtractor \citep{bertin96} to detect and measure the photometry
of sources in each of our reduced images.  Low signal-to-noise (S/N)
borders were masked interactively and detections here were excluded.  We
used a DEBLEND\_MINCONT value of 0.003, DETECT\_MINAREA set to 8 pixels,
and a convolution with a 3-pixel Gaussian filter.  We experimented with
the detection threshold, finding that a value of 1.2$\sigma$ detected
all potential sources in the data.  Deblending is an obvious concern in
close pair studies.  We inspected many deblended sources, comparing them
to their counterparts in the HST $z$-band imaging, and verified that the
deblending algorithm successfully identifies what appears to be separate
galaxies and not sub-components of individual systems.  This problem is
less extreme in our 0\farcs5 seeing MOIRCS imaging---which smooths over
fine sub-structure---than in the higher resolution HST data.
Additionally, we mitigate the potential problem of over-deblending by
setting a minimum to the pair separation ($r_{\rm sep} > 5 h^{-1}$ kpc)
in the analysis that follows, although this has little effect on our
results.

To investigate the completeness and photometric uncertainty of our data,
we inserted numerous fake sources of varying magnitudes into the reduced
images and then recovered them using the same SExtractor parameters that
were applied to real sources.  Fake sources were given Gaussian profiles
with FWHM values similar to point sources.  We define the image depth by
the magnitude corresponding to a recovery rate of 80\%.  This depth
implies a slightly lower detection rate (60--70\%) for real galaxies,
which are more extended \citep[e.g.,][]{conselice07}.  The scatter in
recovered magnitudes gives the photometric uncertainty as a function of
magnitude.

\subsection{ISAAC Near-IR Imaging}

In GOODS-S, $JHK_s$ imaging was obtained by an ESO/GOODS
project\footnote{Program ID: LP168.A-0485} using ISAAC on the VLT at the
ESO Paranal Observatory.  The survey details and data reduction are
described in Retzlaff (in preparation).  The final version 2.0 reduced
and calibrated images are publicly available on the ESO website and, for
the \kband\, are similar in quality to our MOIRCS data in GOODS-N.
Covering GOODS-S required 26 pointings given ISAAC's smaller field of
view (2.5\arcmin\ on a side).  The \kband\ coverage amounts to 160
arcmin$^2$ with $\sim$95\% overlap of the GOODS-S ACS imaging.  Images
were also obtained in the $J$ and $H$ bands.  After excluding low S/N
image borders, small gaps ($< 1$\arcsec) separate a few of the ISAAC
pointings.  This is not ideal for companion searches, but has a
negligible impact on our results.  The \kband\ depth varies from 24.1 to
25.2 (AB) with seeing less than 0\farcs6 on all images and typically
more like 0\farcs5.

As with the MOIRCS data, we used SExtractor to perform photometry on the
ISAAC data in all three filters.  We used a DEBLEND\_MINCONT value of
0.0005, DETECT\_MINAREA set to 8 pixels, and a convolution with a
3-pixel Gaussian filter.  A combined source list was assembled with
duplicates flagged in the same way as above.  The image depth and
photometric uncertainty was estimated using the same procedure of
inserting fake sources.  We note that these data were also used by
\citet{rawat08} in a pair fraction study.

\subsection{Matched Catalogs}

We constructed our \K-selected sample in GOODS-N by combining the
detections in all of the MOIRCS images into a single source list.
Because we designed the field layout so that the pointings overlap, duplicate
sources must be removed from this list.  We did this by searching for
those sources that were within 1\arcsec\ of another source but were
detected on a different MOIRCS image.  The duplicate with the larger
number of SExtractor warning flags (bad pixels, saturation, etc.) or
worse photometric uncertainty was flagged and discarded from the
analysis.  The full GOODS-N \kband\ catalog contains 8112 unique
sources over 164 arcmin$^2$.

We matched this \kband\ source list to a number of publicly available
catalogs in GOODS-N.  Beginning with the HST data, we used the publicly
available v1.1 GOODS-N $z$-selected ACS
catalog\footnote{http://archive.stsci.edu/prepds/goods/}
\citep{giavalisco04}.  The search tolerance was 0\farcs5 for $K_s > 22.8$
 and 0\farcs7 for $K_s < 22.8$ to account for possible
centroiding problems for bright sources.  Where the ACS and MOIRCS data
fully overlap, $\approx$86\% of the \kband\ sources have an ACS
counterpart.  For sources with $K_s < 22.8$, this number increases
to $\approx$98\%.

We then cross referenced our \kband\ sample to several spectroscopic
redshift catalogs.  For the Team Keck Treasury Redshift
Survey\footnote{http://tkserver.keck.hawaii.edu/tksurvey/}
\citep[TKRS,][]{wirth04}, we adjusted the TKRS astrometry by
+0.37\arcsec\ in declination and used a search tolerance of 0\farcs6 for
$K_s > 22.8$ and 0\farcs7 for $K_s < 22.8$.  Of the 4364 TKRS matches we
use the 1485 with secure redshifts designated by a ``zquality'' code of
three or four.  We use similar search criteria to match our source list
to the publicly available compilation of spectroscopic
redshifts\footnote{http://www.ifa.hawaii.edu/~cowie/hhdf/acs.html}
initiated by the survey work in \citet{cowie04}, as well as the $1.4 < z
< 3.0$ spectroscopic survey undertaken by \citet{reddy06}.  We also
cross referenced our catalog with the spectroscopic redshifts obtained
by \citet{treu05}.  Including TKRS, these comparisons provide a total of
2109 secure spectroscopic redshifts for the GOODS-N sample.

Finally, to obtain $U$-band observations in order to improve \photoz\
estimates, we cross referenced our catalog with the deep ground-based
photometry obtained by the Hawaii Hubble Deep Field North
Survey\footnote{http://www.astro.caltech.edu/~capak/hdf}
\citep[HHDF-N,][]{capak04}.  We downloaded the $R$-selected catalogs and
adjusted the HHDF-N astrometry by -0\farcs1 in RA and +0\farcs2 in Dec to
match the ACS-based astrometry used here.  The search criteria were the
same as for TKRS.

In GOODS-S, a source catalog based on the ISAAC data was constructed
with the same methods used to build the MOIRCS catalog in GOODS-N.
Matching to the v1.1 GOODS-S $z$-selected ACS catalog was also carried
out in a similar fashion.  The \K-selected ISAAC catalog was correlated to
source catalogs obtained from the ISAAC $J$ and $H$-bands ($JH$
photometry was not obtained in GOODS-N).  The final
ISAAC \K-selected catalog in GOODS-S contains 9043 unique sources over
160 arcmin$^2$.

As with the MOIRCS-based catalog in GOODS-N, we matched the ISAAC
catalog to several publicly available surveys.  We downloaded Version
1.0 of the VIMOS GOODS/ADP Spectroscopic Survey\footnote{Based on
  observations made with ESO Telescopes at the La Silla and Paranal
  Observatories under Program ID 171.A-3045.}  \citep{popesso08} and
Version 3.0 of the FORS2 GOODS/ADP Spectroscopic
Survey\footnote{Observations were carried out using the Very Large
  Telescope at the ESO Paranal Observatory under Program IDs:
  170.A-0788, 074.A-0709, and 275.A-5060.} \citep[][and references
therein]{vanzella08}, from the ESO/GOODS project
website\footnote{http://www.eso.org/science/goods}.  Matched photometry,
several more spectroscopic redshift surveys, and high quality \photoz\ estimates are compiled in the
GOODS-MUSIC sample \citep{grazian06}.  We matched the VIMOS, FORS2, and
MUSIC catalogs to our ISAAC \K-selected catalog using a search tolerance
of 0\farcs6.  Only ESO redshifts with quality codes of A or B or MUSIC
codes less than or equal to 1 (signifying high confidence in the given
redshift) are used in our analysis.  This provides 1683 secure
spectroscopic redshifts for GOODS-S.  

To summarize, our initial \K-selected catalog, complete to $K_s \approx
23.8$, consists of 8112 sources in GOODS-N and 9043 in GOODS-S for a total
of 17155.  Of these, 14998 have ACS counterparts and 3792 have secure
spectroscopic redshifts.  In \S\ \ref{methods} we build a mass-limited
galaxy sample from these matched catalogs by imposing a magnitude cut of
$K_s < 23.57$.  The resulting sample forms the basis of our close pair
analysis.  Its properties are discussed in \S\ \ref{methods}.

\subsection{Supplemental Photometric Redshifts}\label{redshifts}

Redshifts are a key ingredient in this work.  Not only are they
necessary for examining how the pair fraction evolves with time, but
they are also used to reject chance projections and confirm true physical
pairs.  Unfortunately, spectroscopic redshifts are not always available
and photometric redshifts are needed in both GOODS fields.  

Previous \photoz\ estimates have been made by several authors, though
not typically for \K-selected samples.  In GOODS-N, high quality
estimates are available from the HHDF-N Survey, but these use
ground-based multiwavelength images that are degraded to the worst
seeing---1\farcs3 in the $U$-band---in order to improve matched
photometry which is performed in 3\arcsec\ diameter apertures.  We found
that faint companions with separations $r < 3$\arcsec\ may not be
detected or may have contaminated photometry in images with such poor
seeing \citep{capak04}, and so we do not use the HHDF-N \photoz\
estimates in this paper.   

Instead, for the 4703 sources (out of 6812 within ACS coverage) in GOODS-N without
spectroscopic redshifts, we use the Bayesian Photometric Redshift code
(BPZ) described in \citet{benitez00}.  In GOODS-N, we include the
$U$-band photometry when available from the HHDF-N survey combined with
2\arcsec\ diameter photometry in the ACS $BViz$ and MOIRCS $K_s$ bands.
Comparing to the spectroscopic redshifts available, \photoz\ outliers
(defined by $| z_{\rm spec} - z_{\rm phot}| > 1$) account for 4\% of the
BPZ estimates, with $\sigma_{\Delta z/(1+z)} \approx 0.1$ once outliers
are excluded.  

In GOODS-S, \photozs\ are required for 6636 sources out of the total
8319 sources within the ACS region.  Precise \photozs\ for 5196 were estimated using an ACS $z$-selected
sample combined with careful PSF fitting and were released in the
GOODS-MUSIC compilation.  Excluding outliers, which account for a
few percent, \citet{grazian06} obtain a \photoz\ precision of
$\sigma_{\Delta z/(1+z)} \approx 0.03$.  We use these high-quality \photozs\ whenever
possible in GOODS-S, but still require an additional 1440 \photozs\ to provide
redshifts for full GOODS-S catalog.  We again use BPZ with
$U$-band photometry from the MUSIC compilation combined with the ACS and
$JHK_s$ ISAAC photometry measured in 2\arcsec\ diameters.  Comparing to
spectroscopic redshifts in GOODS-S, we find an outlier fraction of the
BPZ \photozs\ of $\sim$10\% and $\sigma_{\Delta z/(1+z)} \approx 0.11$ with outliers
excluded.  This poorer performance of BPZ in GOODS-S compared
to GOODS-N---despite the addition of the $J$ and $H$ bands---seems to
result from the shallower $U$-band photometry in GOODS-S as well as a
discrepancy in the ISAAC filter curves\footnote{The ISAAC filter
  transmission curves released on the ISAAC instrument website (which
  give $K_{\rm AB} - K_{\rm Vega} = 1.761$) do not match the latest
  calibrations in the final Version 2.0 release.  The documentation
  reports that the throughput has been remeasured for Version 2.0, giving $K_{\rm AB} -
  K_{\rm Vega} = 1.895$.  Updated ISAAC transmission curves---an input
  to BPZ---have not been made available, however.}

In both fields comparisons to spectroscopic redshifts with
$z \gtrsim 1.5$ revealed additional catastrophic \photoz\ failures (with
incorrect estimates typically assigned to $z_{\rm BPZ} \lesssim 0.5$).
Without redshift comparisons at $z \gtrsim 1.5$, we would have overestimated
our \photoz\ precision.  It is worth emphasizing that judgments on the
quality of \photoz\ estimates from any survey are limited by
the nature of the spectroscopic comparison set that is available.

\section{Galaxy Properties}\label{properties}

The deep \K-selected catalogs described above provide a powerful dataset
for our galaxy pair analysis.  Before describing the mass-limited sample
we draw from this dataset and the methodology we use to study pairs, we
present our techniques for estimating three key physical galaxy
properties that allow us to characterize the role of merging among
different galaxy types.

\subsection{Stellar Masses}

The first of these properties is the stellar mass ($M_*$), which is
estimated using the Bayesian code described in \citet{bundy06}.  SED
fitting is performed using the 2\arcsec\ diameter $BVizK_s$ photometry
and the best available redshift.  The observed SED of each galaxy is
compared to a grid of 13440 models from the \citet{bruzual03} population
synthesis code that span a range of metallicities, star formation
histories (parametrized as exponentials), ages, and dust content.  No
bursts are included in our models and only those models with ages
(roughly) less than the cosmic age at each redshift are considered.  We
use a Chabrier IMF \citep{chabrier03} and assume a Hubble constant of 70
km s$^{-1}$ Mpc$^{-1}$.

At each grid point, the \kband\ $M_*/L_K$ ratios, inferred $M_*$, and
probability that the model matches the observed SED is stored.  This
probability is marginalized over the grid, giving the stellar mass
probability distribution, the median of which is taken as the final
estimate of $M_*$.  The width of the distribution provides the
uncertainty which is typically 0.1--0.2 dex.  This is added in
quadrature to the \kband\ magnitude uncertainty to determine the final
error on $M_*$.  Stellar mass estimates for galaxies with \photozs\ also
suffer from the uncertainty in luminosity distance introduced by the
\photoz\ error and the possibility of catastrophically wrong
redshift information \citep{bundy05, kitzbichler07}.  We will take both into
account in the analysis that follows.

More broadly, any stellar mass estimate suffers potential systematic
errors from uncertainties inherent in stellar population modeling and
various required assumptions, such as the form of the IMF.  Several
papers have stressed the importance of treating thermally pulsating
asymptotic giant branch stars \citep[TP-AGBs, e.g.,][]{maraston06} an
element that is missing in the \citet{bruzual03} models.  The recent and
thorough investigation of population synthesis modeling by
\citet{conroy08}, however, argues that $M/L$ ratios estimated from
\citet{bruzual03} are largely resistant to the uncertain contribution
from TP-AGBs as well as other limitations of current models.  Still, it
is important to recognize that $M_*$ estimates may be affected by
unrecognized systematic uncertainties at the 0.1--0.2 dex level.

\subsection{Rest-Frame Color}

We use the inferred rest-frame color to roughly split our sample into
blue, star-forming galaxies and red, quiescent ones \citep[see][]{willmer06}.  We use the ACS
rest-frame $(B - i)$ color, with $k$-corrections determined from the SED
fit performed by our stellar mass estimator.  For each galaxy, the color
of every model galaxy in the grid is weighted by the probability that
the model matches the observed SED.  The weighted colors are then summed
over the grid to derive the probability distribution as a function of
$(B-i)$ color.  As with $M_*$ we take the median of the distribution as the final
color estimate.

The $(B-i)$ distribution across the sample is bimodal, as expected.  We
define red galaxies to be those redder than $B-i=1.3$ (AB units) in this
filter system.  This threshold results in very similar red/blue
distributions and number densities as those reported in
\citet{willmer06} and \citet{bundy06} using data from the DEEP2 Redshift
Survey.  We note however that while a restframe color partition is easy
to implement, it is still a blunt method for discriminating star-forming
from passive galaxies.  For example, contamination from dusty
star-forming galaxies on the red sequence is at least $\sim$10\%
\citep[see][]{yan06} and a comparison to morphological types (see below)
for a subset of galaxies with $0.6 < z < 0.8$ and $M_{\odot} > 10^{10}$
shows that $\sim$12\% of blue galaxies are classified as spheroidal
while $\sim$24\% of red galaxies are classified as disk-like.

\subsection{Morphology}

Galaxy morphology is the final property we consider and is important
because morphological evolution may be driven in part by merging.
Morphological classification in the GOODS fields was presented in
\citet{bundy05} and is publicly
available\footnote{http://www.astro.caltech.edu/GOODS\_morphs}.  The
classification was carried out visually to a magnitude limit of $z_{\rm
  AB} = 22.5$ which increases the effective mass limit of subsamples
split by morphology, as we discuss below.  Each source was assigned one
of the following morphological types: -2=star, -1=compact, 0=E, 1=E/S0,
2=S0, 3=Sab, 4=S, 5=Scd, 6=Irr, 7=Unclass, 8=Merger, 9=Fault.  As in
\citet{bundy05}, we group these definitions into the broader categories
of spheroidals (types 0--2), disks (types 3--5), and irregulars (types
6--8).  Visual morphologies are available for 1369 sources in GOODS-N
and 1200 in GOODS-S.  

\section{Methods for Counting Galaxy Pairs}\label{methods}

In any study of galaxy mergers, it is important to distinguish major
from minor mergers, since major mergers---typically defined to have
mass ratios between 1:4 and 1:1---have the potential to radically affect
its morphology and star formation \citep[e.g.,][]{naab03}.  With many previous studies lacking
near-IR data, however, pair samples comprised an unknown mix of mass
ratios.  In fact, pairs defined using an optical magnitude difference
may lead to inflated pair counts from triggered SF \citep{bundy04}, an
effect that may worsen at higher redshifts where the global SFR
increases.  Thus an important goal of this work is to use our \K-selected
sample to select only those pairs with stellar mass ratios consistent with major
mergers.  

We implement this requirement by selecting pairs with a magnitude
difference\footnote{Recognizing the assumptions inherent in estimating
  $M_*$, our decision to use the {\em observed} magnitude difference as
  opposed to the {\em inferred} mass difference is designed to
  facilitate future comparisons to our results.} of $\Delta K_s \le 1.5$.
Using pairs where mass estimates are available for both members, we find
that this magnitude criterion results in mass ratios
$M_{companion}/M_{host} \ge 1/4$ for $\sim$80\% of the pairs.  This number is
greater than $1/6$ for 90\%.  Pairs are only included if found within a
projected annulus of $5 h^{-1} {\rm ~kpc} < r_{\rm sep} < 20 h^{-1} {\rm
  ~kpc}$ determined by the redshift of the primary or ``host'' galaxy,
given our assumed cosmology ($H_0$=70 km s$^{-1}$ Mpc$^{-1}$,
$\Omega_M$=0.3 and $\Omega_{\Lambda}$=0.7).  We avoid double counting by
identifying only those companions that are fainter than their host
galaxy.  In order to infer the true physical pair fraction, we use two
methods to remove chance projections from the raw pair counts defined
above.  The first uses a correction based solely on the sky density of
background sources while the second employs redshift information to
identify pairs in 3D space.

Before describing these methods in detail, we begin with a description
of our mass-limited sample of host galaxies.  It is important that the
sample be clearly defined because the definition of the pair fraction,
\fpair, is determined with respect to such a sample.  With our
definition, the number of paired systems is $N_{\rm pair} = f_{\rm pair}
N_{\rm gal}$, where $N_{\rm gal}$ is the number of host galaxies.  Note
that this definition differs from that of other authors in that \fpair\ is not
the fraction of galaxies located in pairs (often termed $N_c$) but
instead reflects the potential number of mergers within a sample
($f_{\rm pair} \approx N_c/2$).  Finally, because our sample allows us to
determine not only the fraction, but the comoving number density of
close pairs, we also describe our technique for estimating that quantity
in this section.  Converting the observed pair fractions into merger
rates requires additional and uncertain assumptions, and we save this
topic for \S\ \ref{discussion}.

\subsection{The Host and Companion Galaxy Sample}

One of the primary advantages of our dataset is that both host and
companion galaxies are drawn from a well-understood, \K-selected catalog
that can be characterized in terms of mass completeness.  Based on the
different depths of our near-IR images, we optimize our sample by choosing a
\kband\ magnitude limit of $K_s = 23.57$.  All companions and host
galaxies must be brighter than this limit.  Because companions are
defined to be no more than 1.5 magnitudes fainter in the \kband\ than
their corresponding host galaxy, this effectively means that our host
sample is limited to $K_s < 22.07$.

Throughout this paper we will adopt three redshift intervals: $0.4 < z <
0.7$, $0.7 < z < 0.9$, and $0.9 < z < 1.4$.  Given the \kband\ limit
imposed, the host sample is complete in each redshift bin for masses
greater than $\log M_*/M_{\odot} =$ 9.6, 10.1, and 10.4.  The
corresponding completeness limits for potential companion galaxies at
these redshifts are $\log M_*/M_{\odot} =$ 9.0, 9.5, and 9.8.  When
morphological samples are considered, an additional $z$-band magnitude
limit of $z_{\rm AB} < 22.5$ is effectively imposed since reliable
morphologies are difficult to obtain at fainter magnitudes.  This
increases the mass completeness limits of host galaxies to $\log M_*/M_{\odot} =$ 10.3,
10.6, and 11.2.  Below these values, some hosts may not be included
in the sample.  We highlight the impact of these limits on our
results as we present them.  These stellar mass limits are estimated by
considering the detection efficiency of redshifted SED templates of
burst-like stellar populations with a formation redshift of $z_{\rm
  form} = 5$.  Because of their high $M_*/L$ ratios, these models
provide conservative and accurate mass limit estimates.

Since companions are defined by being near their host galaxy, the field
geometry of the sample is also important.  We define contiguous survey
regions within both GOODS fields where the \kband\ detection limit is
greater than 23.57 and the ACS tiles maximally overlap (we exclude the
triangular, low-S/N perimeters of the GOODS footprint).  All of the
ISAAC images in GOODS-S meet this requirement, giving a search area
of 147 arcmin$^2$.  We exclude three of the 16 MOIRCS images in
GOODS-N because they are shallower than $K_s = 23.57$.  This gives a
final area of 139 arcmin$^2$ in the northern field, or 286 arcmin$^2$
with both fields combined.  Only host galaxies separated by a distance
greater than $r_{\rm max}$ from the border of the defined region are
included (this removes $\sim$3\% from the sample).  Stars, as determined visually in the morphology catalog of
\citet{bundy05} and also as identified spectroscopically, are removed.
The final host sample includes 2994 primary galaxies, more than 60\% of
which have spectroscopic redshifts.

Unfortunately, spectroscopic redshifts are not always available for both
the host galaxy and potential companions.  We must therefore make
statistical corrections that account for the contamination of the true
{\em physical} pair fraction from close, line-of-sight projections.  We
describe the two approaches we adopt for this correction below.

\subsection{Method I: Projected Field Correction}\label{methodI}

Our first method for correcting the raw pair counts is common in past studies at high redshift
\citep[e.g.,][]{le-fevre00} and assumes no redshift information for
fainter companions is available.  We require only that potential
companions be detected in the \kband\ and that they satisfy the criteria
described above.  The field contamination for each host galaxy is
measured according to the \kband\ source density (i.e., number counts)
observed in the defined survey region in the magnitude interval $K_{\rm
  host} < K_{\rm companion} < K_{\rm host} + \Delta K$, with $\Delta K = 1.5$.
Because of cosmic variance, it is important to make this measurement
separately for the host sample in GOODS-N and GOODS-S.  For a given
host galaxy, the measured sky density of possible contaminants is
multiplied by the area of the corresponding search annulus to determine
the contamination rate per galaxy.  This is then summed across the
potential host sample to estimate the total number of field
contaminants, $N_{\rm corr}$.  We estimate the number of true physical
pairs by subtracting $N_{\rm corr}$ from the total number of potential
companions fulfilling the search criteria described above.  As discussed
in detail in \S \ref{results}, this method identifies 242 pairs within
our sample, $\sim$160 of which are expected to by chance projections.
Uncertainties are determined by adding the Poisson error from $N_{\rm
  corr}$ and from the total raw pair counts in quadrature.  We expect
an additional contribution from cosmic variance in the relatively small
GOODS fields.

\subsection{Method II: Probable Redshift Confirmation}\label{methodII}

The vast majority of faint companions identified in the \kband\ catalog
are also detected in the ACS and ground-based optical imaging which
means that photometric redshifts can be estimated for these sources.  A
sizable fraction of companions ($\sim$30\%) also have spectroscopic
redshifts.  In our second pair counting method, we use the redshift
information available for faint near-by neighbors to find those that are
likely to be physically associated with their brighter host.  This
amounts to an additional redshift constraint on the identification of
companions.  For both the potential hosts and companions we define a
redshift uncertainty of $\sigma_z = 0.002$ for galaxies with
spectroscopic redshifts, $\sigma_z = 0.08(1+z)$ for galaxies with
photometric redshifts\footnote{The value of $0.08(1+z)$ is somewhat
  arbitrary and represents a compromise between the BPZ \photoz\
  uncertainty ($\sim$0.1) and the MUSIC \photoz\ uncertainty
  ($\sim$0.03).}, and an unbounded $\sigma_z$ for the remaining few
galaxies without optical detections and therefore with no redshift
information.  When the redshift difference satisfies $\Delta z^2 <
\sigma_{z,host}^2 + \sigma_{z,companion}^2$, the pair is
selected\footnote{For the pairs in our sample identified by method II,
  $\sim$26\% feature spectroscopic redshifts for both members,
  $\sim$39\% require at least one \photoz\, $\sim$28\% require \photozs\
  for both members, and $\sim$7\% are without redshift information for
  the fainter companion} with this method.

Because of the large redshift uncertainty, some of the host-companion
pairs defined by photometric redshifts will be close in projection but
not physically associated.  As in our first pair counting method, a
contamination correction is needed although it will of course be smaller
since most contaminants are already excluded by their discrepant
redshifts.  As in \citet{kartaltepe07}, we determine this correction
factor by randomizing the coordinate positions of all the sources in our
catalog (but retaining all redshift and magnitude information).  Any
companions identified in the randomized data are clearly chance
superpositions.  Thus, by repeating this exercise for 100 randomized
datasets, we can estimate the average contamination rate, its
uncertainty, and the dependence of these quantities on $M_*$ and pair
type (see below).  Note that in the limit that no redshift information
is available for faint neighbors this approach is equivalent to the
first method described above.  However, for surveys that contain dense
environments like galaxy clusters (GOODS is too small to sample such
environments), this technique underestimates the needed correction
because it does not account for the increased rate of chance projections
in dense regions.

Finally, we note that the possibility of catastrophic errors in \photoz\
estimates will cause some true physical companions to be disregarded as
chance projections.  We can determine the importance of this effect by
examining the quality of our various \photoz\ estimates (\S\
\ref{redshifts}) and estimating the likely number of catastrophic
\photoz\ failures that cause us to underestimate the true number of
pairs in method II.  Based on this analysis we correct all \fpair\
values derived from method II upwards by 1\%.  Using method II, we find
89 pairs, $\sim$ 41 of which are expected to be chance projections.

\subsection{Estimating Comoving Number Densities}\label{method_MF}

One of the important quantities we wish to determine in this work is the
comoving volumetric merger rate as a function of the stellar mass of the
host galaxy.  As we discuss in \S\ \ref{discussion}, this ``merger rate
mass function'' provides a powerful tool for understanding the role of
galaxy mergers in driving mass assembly as well as evolution in the
numbers of different populations such as ellipticals and red-sequence
galaxies.

We  estimate such mass functions using the simple $V_{max}$
technique \citep{schmidt68}.  We weight
host galaxies and associated pairs by the maximum volume in which they would
be detected within the \kband\ limit in a given redshift interval.  In practice, our
magnitude limits impose restrictions only in the highest redshift bin
($0.9 < z < 1.4$).  For each host galaxy $i$ in the redshift interval $j$, the
value of $V^i_{max}$ is given by the minimum redshift at which the
galaxy would drop out of the sample,

\begin{equation}
V^i_{max} = \int_{z_{low}}^{z_{high}} d \Omega \frac {dV}{dz} dz
\end{equation}

\noindent where $d \Omega$ is the solid angle subtended by the survey area, and $dV/dz$ is the comoving
volume element.  The redshift limits are given as,

\begin{equation}
z_{high} = {\rm min}(z^j_{max}, z^j_{K_{lim}})
\end{equation}
\begin{equation}
z_{low} = z^j_{min}
\end{equation}

\noindent where the redshift interval, $j$, is defined by $[z^j_{min},
z^j_{max}]$ and $z^j_{K_{lim}}$ refers to the redshift at which the
galaxy would still be detected below the \kband\ limit for that
particular redshift interval.  We use the best-fit SED template as
determined by the stellar mass estimator to calculate $z^j_{K_{lim}}$,
thereby accounting for the $k$-corrections necessary to compute
$V_{max}$ values (no evolutionary correction is applied).

\section{Results}\label{results}

\subsection{The Stellar Mass Dependent Pair Fraction}

As discussed in \S\ \ref{intro}, the hierarchical framework that
underlies current galaxy formation models argues that the mass scale on
which galaxies assemble increases with time.  Observations of star
formation quenching and morphological evolution show that these
processes are set by a mass scale that decreases with time
(downsizing).  Meanwhile, galaxy mergers are thought to be key
drivers of both hierarchical assembly and the evolution of SF and galaxy
morphology.  So, understanding the mass dependence of merging provides a
critical test of the role mergers play in galaxy mass assembly
and evolution.

\begin{figure}
\epsscale{1.1}
\plotone{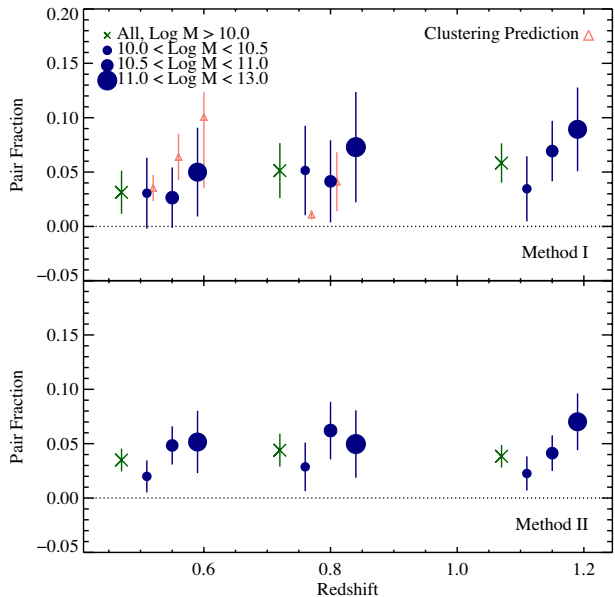}
\caption{Contamination-corrected fraction of paired systems with $5 h^{-1} {\rm ~kpc} < r_{\rm sep} < 20 h^{-1} {\rm
  ~kpc}$ among
  galaxies in three redshift intervals.  The top panel shows results
  determined using method I, the bottom panel method II.
  Abscissa values have been spaced to keep the data points from
  overlapping.  The X symbols denote the full sample of primary (host) galaxies while the size of
  the circular symbols indicates the result of dividing these
  into stellar mass bins, as labeled.  Predicted values of \fpair\ based on the 2-point
  correlation function (see \S\ \ref{clustering}) of \citet{zehavi05}
  (using SDSS) are overplotted in the
  low-$z$ bin of the top panel while the mid-$z$ bin shows predictions based on
  \citet{coil06} (using DEEP2).  The mass dependence of the clustering predictions is
  based roughly on the luminosity range probed by the clustering analyses.
  \label{fig:frac_mass}}
\end{figure}

We begin our investigation of how galaxy mergers depend on mass by
presenting the pair fraction measured using the two methods described in
the previous section.  With method I, in which no redshift information
for companion galaxies is used, 242 pairs are found, $\sim$160 of which
are expected to be chance projections.  The method I pair fractions,
partitioned by mass and redshift, are plotted in the top panel of Figure
\ref{fig:frac_mass}.  With method II, which takes companion redshifts
into account, 89 pairs are found, $\sim$41 of which are expected
projections.  The resulting method II pair fractions are plotted in the
bottom panel of \ref{fig:frac_mass}.  In both cases, pair fractions are
plotted for three redshift intervals: $0.4 < z < 0.7$, $0.7 < z < 0.9$,
and $0.9 < z < 1.4$.  The full sample (X symbols) as well as the
dependence on the mass of the host, as indicated by the differently
sized circular symbols, is shown.  All of the plotted points are
complete in terms of stellar mass except for the highest redshift $10 <
\log M_*/$\msun$ < 10.5$ bin which is $\sim$80\% complete as a result of
the \kband\ limit.  The results from both methods are listed in Table
\ref{table:fpair}.

We first note that the two panels in Figure \ref{fig:frac_mass} are similar, reinforcing the utility of the
two methods.  Both plots show that roughly 2--5\% of galaxies in a
mass-limited sample host fainter companions.  The value of \fpair\
depends on the maximum allowed pair separation, $r_{\rm max}$.  The
average value of \fpair\ increases by an amount of 0.03 for $r_{\rm max} = 25 h^{-1} {\rm ~kpc}$
compared to $r_{\rm max} = 15 h^{-1} {\rm ~kpc}$.  This difference does
not impact the derived merger rate because the expected merger
efficiency decreases with larger $r_{\rm max}$
\citep[e.g.,][]{patton08}.  We use $5 h^{-1} {\rm ~kpc} < r_{\rm sep} <
20 h^{-1} {\rm ~kpc}$ in what follows.

The key result in Figure \ref{fig:frac_mass} is the increase in \fpair\
among more massive host galaxies.  Hints of this trend were observed by
\citet{xu04} in their $K$-selected study of close pairs at $\langle z
\rangle = 0.03$.  We confirm it here with moderate significance to $z
\gtrsim 1$ using both pair count correction methods.  We verified that
these results are still apparent for different values of $r_{\rm sep}$
including $r_{\rm sep} = 15 h^{-1}$kpc and $r_{\rm sep} = 25 h^{-1}$kpc.
The mass dependence is also apparent in each GOODS field separately,
suggesting that this trend is not a result of cosmic variance.

As we discuss in \S\ \ref{discussion}, some of the increase at higher
masses may reflect the stronger clustering of such galaxies.  In that
section, we return to the important question regarding the extent to
which the mass dependence observed for the pair fraction translates into
a mass dependence for the merger rate.  Still, the fact that the lowest
mass bins in Figure \ref{fig:frac_mass} are nearly consistent with a zero pair
fraction implies that the merger rate inferred from this and other
studies is dominated by higher mass galaxies.  We show in \S\
\ref{vol_merge} that this leads to the result that major mergers are
unlikely to be the sole mechanism behind the formation of spheroidal
galaxies and the red sequence.  At the same time, the higher merger
rates implied for massive galaxies shows that they are continuing to
assemble after $z \sim 1$.  This is an important piece of direct
evidence for hierarchical growth which we seek to quantify in \S\
\ref{vol_merge}.

Finally, we comment on the possibility that \fpair\ evolves with
redshift.  Averaging over the full mass range, $M_* > 10^{10}
M_{\odot}$, we find no statistically significant evolution.  With
$f_{\rm pair} \propto (1+z)^{m}$, we measure $m = 1.6 \pm 1.6$ using
method I and $m = 0.3 \pm 1.4$ using method II.

\subsection{The Dependence on Host Galaxy Color and Morphology}

In addition to their role in mass assembly and their potential to affect
the morphology and SFR of galaxies, mergers have also been invoked as a
means to build the massive end of the red sequence.  Successive dry
mergers between early-type galaxies at lower masses can increase the
numbers at higher masses \citep[e.g.,][]{van-dokkum05a, faber07}.  In
this scenario, mergers must clearly be common within the red-sequence.
We can begin to test this by characterizing the dependence of \fpair\ on host
galaxy type.  Our goal is not only to count the frequency of dry
mergers, but also the remaining fraction of wet mergers since only
these events can be responsible for morphological transformation or star
formation quenching.

\begin{figure}
\epsscale{1.2}
\plotone{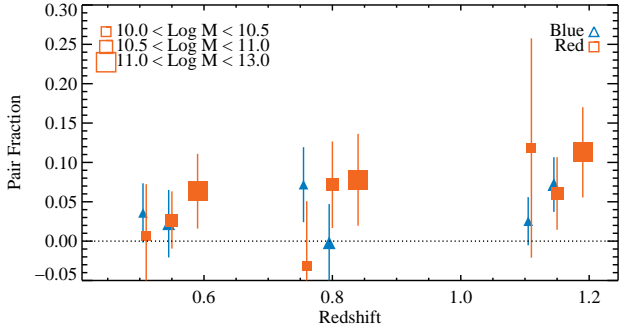}
\caption{Color and mass dependence of the field-corrected pair fraction,
  \fpair\, using method I.  The host sample has been divided into ``blue'' (blue
  triangles) and ``red'' (red squares)
  based on rest-frame color.  Symbol size indicates stellar mass as in
  Figure \ref{fig:frac_mass}.  Bins corresponding to the most
  massive blue galaxies have not been plotted because they contain fewer than 10 hosts.
  \label{fig:pair_frac_col}}
\end{figure}

Both pair counting methods offer insight on the question of how the pair
fraction varies with galaxy type.  Since the first method relies on an
entirely statistical field correction, we can only use it to study
trends with host galaxy type.  The properties of the companion and host
together---the wet and dry merger frequency, for example---are
accessible with method II and discussed below.  Method I allows us to divide the host
sample by both mass and type.  Figure \ref{fig:pair_frac_col} shows the
result when host galaxy type is defined using the color bimodality to
separate star-forming from passive systems (also see Table
\ref{table:fpair}).  Bins corresponding to the most massive blue
galaxies have not been plotted because they contain fewer than 10 hosts.

While the uncertainties have increased compared to Figure
\ref{fig:frac_mass}, there is weak evidence that the mass-dependent
trend observed in Figure \ref{fig:frac_mass} is reflected in the
star-forming properties of the host galaxies, with higher pair fractions
found for quenched, red hosts, in large part because such galaxies
dominate at the highest masses.  This suggests a role for dry mergers in
building the most massive early-type galaxies.  However, a second
interpretation discussed further in \S\ \ref{discussion} is that Figure
\ref{fig:pair_frac_col} may reveal the greater degree of clustering
among massive red galaxies.  Finally, we note that the number of red
hosts with companions in the lowest mass bin is consistent with zero.
If confirmed with future studies, this observation would indicate that
major mergers among red systems with $M_* \lesssim 10^{10}$\msun\ cannot
contribute significantly to the increasing abundance of red sequence
galaxies at higher masses.  Transforming blue galaxies into red objects
via quenched star formation would therefore appear to be a more
important mechanism for red sequence growth at $M_* \lesssim
10^{11}$\msun.

\begin{figure}
\epsscale{1.2}
\plotone{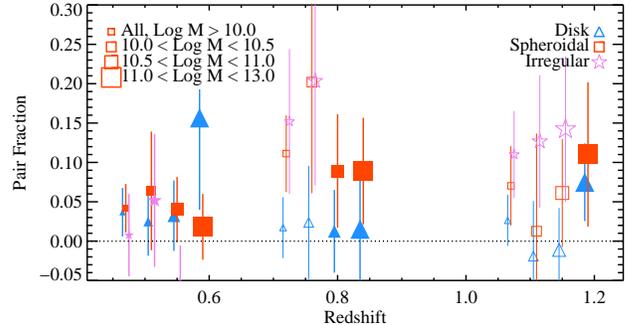}
\caption{Morphological dependence of the field-corrected pair fraction
  using method I.
  The host sample has been restricted to $z_{\rm AB} < 22.5$ and visually
  classified by ACS morphology into disks (blue
  triangles) and spheroidals (red squares) and irregulars (magenta stars).  Symbol size indicates stellar mass as in
  Figure \ref{fig:frac_mass} with the smallest samples denoting the
  full mass range.  Data points corresponding to bins
  that are significantly incomplete in stellar mass are indicated by lighter colors
  and open symbols.
  \label{fig:pair_frac_morph}}
\end{figure}

In Figure \ref{fig:pair_frac_morph} we explore the relationship between
pair fraction and galaxy morphology.  Here the host sample has been
restricted to $z_{\rm AB} < 22.5$ which makes some mass bins incomplete
(although we note we are measuring fractions not absolute quantities).
Bins that are less than $\sim$80\% complete incomplete are indicated by
lighter colors and open symbols and will tend to have an artificially
lower number of redder galaxies.  Given the large error bars, any trends
observed here will require confirmation from larger studies.  Still,
there is a hint that spheroidals and irregulars host more companions
than disk galaxies.  Spheroidals, which are expected to be mostly red
and quiescent, also tend to exhibit higher pair fractions among
high-mass galaxies with $\log M_*/$\msun$ > 10.5$ (although this is not
the case at $z \sim 0.5$).

One intriguing and more robust result of Figure
\ref{fig:pair_frac_morph} is the appearance of a high fraction of
irregular hosts at intermediate masses.  In some ways, this is
unexpected because irregulars represent only 10--20\% of the galaxy
population in our redshift range with $\log{M_*/M_{\odot}} \gtrsim 10$
\citep{brinchmann00, bundy05, pannella06}.  They should therefore be
less likely to host companions, if pairs were randomly distributed among
all galaxy types.  Even in incomplete bins which are biased against red
spheroidals, host galaxies are more likely to have irregular
morphologies than disk-like ones.  Because we ignore companions with
$r_{\rm sep} < 5 h^{-1}$kpc, the prevalence of irregular hosts
tentatively suggests that even more distant companions can imprint
significant morphological disturbances on their host galaxy, perhaps
after executing a first pass \citep[e.g.,][]{lotz08a}.  A similar
conclusion was reached by \citet{li08} using SDSS.  They also find an
enhancement of SF among close pairs which could lead to morphological
peculiarities.  It stands to reason that not all irregular systems are
in the last stages of a significant merger, which may in part explain
why merger rates based on disturbed morphologies are typically higher
than those inferred from pair counts \citep[e.g.,][]{conselice03,
  lotz08}.

Taken together, the type dependent pair fractions we have presented
suggest that galaxies with higher masses, redder colors, and early-type
morphologies tend to host companions more frequently than their lower
mass, star-forming, and disk-like counterparts.  This lends support to
the concept of dry mergers as a mechanism for red sequence growth,
especially at the highest masses, a point we return to in \S\
\ref{mass_assembly}.  However, dry mergers do not appear to be important
for building the red sequence at $M_* \sim 10^{10}$\msun\ because the
number of red hosts at these masses is nearly zero.  At the same time,
the enhanced pair fraction for more massive red and spheroidal galaxies
may simply reflect the denser environments of these galaxies.  The
question of whether the derived merger rates are also enhanced in such
environments is discussed in \S\ \ref{discussion}.

\subsection{The Frequency of ``Dry'' and ``Wet'' Pairs}\label{dry_wet_pairs}

\begin{figure}
\epsscale{1.1}
\plotone{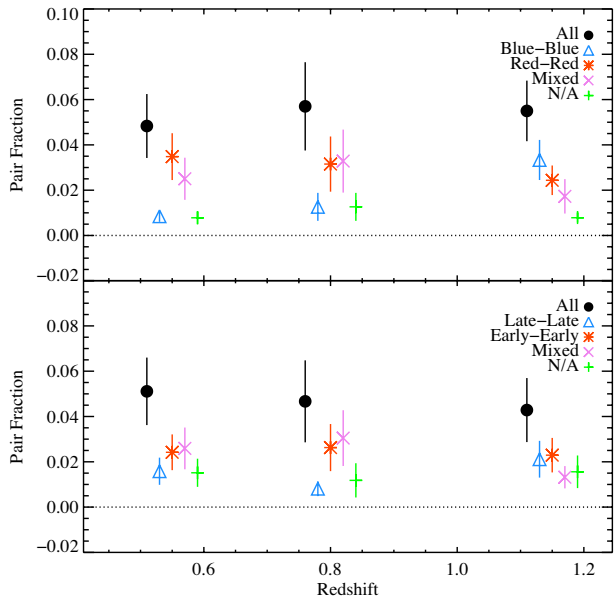}
\caption{Fraction of dry, wet, and mixed pairs as defined by both color
  (top panel) and morphology (bottom panel). for host
  galaxies with $\log M_*/M_{\odot} > 10.5$.  Pairs are confirmed using spectroscopic and
  photometric redshifts, and a correction for contamination and
  catastrophic \photozs\ has
  been applied (method II).   Pairs including secondaries without a
  redshift are listed as ``N/A.''  In the bottom panel, early-type refers to galaxies with
  spheroidal morphologies (E/S0/Sab) while late-type includes both
  disk-like and morphologically irregular galaxies.   An additional
  magnitude cut of $z_{\rm AB}
  < 22.5$ applies to the bottom panel.  
  \label{fig:zpp}}
\end{figure}

In the top panel of Figure \ref{fig:zpp} we use method II combined with the
rest-frame color of pair members to distinguish the fraction of pairs
expected to result in dry and wet mergers (also see Table
\ref{table:dry_wet}).  Red-red pairs are identified as dry merger
candidates.  Blue-blue and mixed pairs indicate the presence of cold gas
reservoirs and therefore represent potential wet mergers.  For the
few secondary (companion) galaxies without a redshift, the pair type
cannot be determined and is labeled as ``N/A.''  Note that a
statistical correction as described in \S\ \ref{methodII} has been
subtracted from each category so that negative values are possible in
some cases.

An interesting evolutionary signal appears in Figure \ref{fig:zpp}.
First, it appears that dry, red-red pairs are less frequent than their
wet counterparts (the sum of blue-blue and mixed) in the highest
redshift bin but become increasingly important with cosmic time,
eventually dominating the low redshift bin.  At the same time the number
of blue-blue pairs decreases to zero.  These trends are significant at
the 1-$\sigma$ level and may simply reflect the changing nature of the
galaxy population with masses $\log M_*/M_{\odot} > 10.5$ which is
increasingly dominated by red galaxies at late times.  If our sample
size allowed us to consider only high mass galaxies (e.g., $\log
(M_*/M_{\odot}) > 11$) we might expect dry pairs among such galaxies to
show an even earlier increase since this population evolves into
quenched red galaxies first \citep{bundy06, borch06, bell07}.  However, it is important
to emphasize that the dry pair fraction is consistent with what is
expected given the makeup of galaxies from which the pairs are drawn.

A similar conclusion was reached by \citet{lin08} who studied the
fractions of blue-blue, red-red, and mixed systems among kinematic pairs
identified in the DEEP2 survey.  They also found a higher number of
blue-blue pairs at $z \sim 1$ compared to the present day combined with
an increasing number of red-red and, to a lesser extent, mixed systems.
While that work benefits from spectroscopic redshifts for all pair
members, our analysis has the advantage of being mass-limited and
capable of selecting only major pairs.  Furthermore, it does not require
weighting to account for biases in the spectroscopic target selection.
Given these very different but complementary methods, it is encouraging
to see agreement both in the derived type-dependent pair fraction (note
the different definitions of pair fraction: $f_{\rm pair} = 0.5N_c$) and
its evolution with redshift.

Our data also allow us to examine the frequency of
pairs defined by the morphological type of both members.  This is shown
in the bottom panel of Figure \ref{fig:zpp} where it is important to remember that in
the highest redshift bin, 30-40\% of red galaxies with $\log
(M_*/M_{\odot}) > 10.5$ will be missed because of the $z_{\rm AB} <
22.5$ morphology cut.  In this panel we label disk-like and irregular
galaxies as ``late-type'' and spheroidals as ``early-type.''  We see no
statistically significant trends, although there is a hint that late-late pairs are on
par with early-early and mixed pairs in the highest redshift bin, but
become subdominant in the mid- and low-z bins.  Although larger datasets will
be needed to confirm this, such an effect would be expected as the
fraction of late-type galaxies declines with time.

As we discuss further in \S\ \ref{discussion}, Figure \ref{fig:zpp} has
important implications on the role of wet mergers in transforming blue
(disk) galaxies into red (spheroidal) galaxies.  Both indicate that the
implied number of such wet mergers is a factor of $\sim$2 less than the
total merger rate.  These figures also show that the rate of dry merging
is not enhanced beyond what is expected given the makeup of galaxies in
a mass-limited sample.  In \S\ \ref{discussion} we show explicitly that
neither morphological transformations occurring after wet mergers nor
mass buildup caused by dry mergers appears sufficient to explain the
increasing number of intermediate to very massive early-type galaxies.


\setcounter{table}{1}
\begin{deluxetable}{lcccccc}
\tablecaption{Method II: Dry, Wet, and Mixed Mergers}
\tabletypesize{\footnotesize}
\tablewidth{0pt}
\tablecolumns{7}
\tablehead{
\colhead{Sample} & \colhead{$z$} & \colhead{} & \colhead{$N_P$} & \colhead{$N_{corr}$} & \colhead{$N_{gal}$} & \colhead{\fpair\ (\%)} \\
\cline{1-7} \\
\multicolumn{7}{c}{Color, $\log M_*/M_{\odot} > 10.5$}
}

\startdata
Blue-Blue & 0.4--0.7  & {} & 4 & 2.3 & 514 & 1.3 $\pm$ 0.5 \\
Blue-Blue & 0.7--0.9  & {} & 7 & 3.3 & 380 & 2.0 $\pm$ 0.8 \\
Blue-Blue & 0.9--1.4  & {} & 24 & 9.4 & 793 & 2.8 $\pm$ 0.7 \\
Red-Red & 0.4--0.7  & {} & 7 & 0.7 & 514 & 2.2 $\pm$ 0.5 \\
Red-Red & 0.7--0.9  & {} & 5 & 0.9 & 380 & 2.1 $\pm$ 0.6 \\
Red-Red & 0.9--1.4  & {} & 8 & 1.3 & 793 & 1.8 $\pm$ 0.4 \\
Mixed & 0.4--0.7  & {} & 8 & 2.6 & 514 & 2.0 $\pm$ 0.6 \\
Mixed & 0.7--0.9  & {} & 10 & 3.5 & 380 & 2.7 $\pm$ 1.0 \\
Mixed & 0.9--1.4  & {} & 10 & 5.8 & 793 & 1.5 $\pm$ 0.5 \\
N/A & 0.4--0.7  & {} & 0 & 1.1 & 514 & 0.8 $\pm$ 0.2 \\
N/A & 0.7--0.9  & {} & 1 & 1.0 & 380 & 1.0 $\pm$ 0.4 \\
N/A & 0.9--1.4  & {} & 1 & 1.2 & 793 & 1.0 $\pm$ 0.2 \\

\cutinhead{Morphology, $\log M_*/M_{\odot} > 10.5$}
Late-Late & 0.4--0.7  & {} & 2 & 1.5 & 482 & 1.1 $\pm$ 0.4 \\
Late-Late & 0.7--0.9  & {} & 0 & 0.4 & 261 & 0.8 $\pm$ 0.3 \\
Late-Late & 0.9--1.4  & {} & 3 & 0.5 & 322 & 1.8 $\pm$ 0.6 \\
Early-Early & 0.4--0.7  & {} & 4 & 0.4 & 482 & 1.8 $\pm$ 0.4 \\
Early-Early & 0.7--0.9  & {} & 3 & 0.2 & 261 & 2.1 $\pm$ 0.7 \\
Early-Early & 0.9--1.4  & {} & 3 & 0.0 & 322 & 1.9 $\pm$ 0.5 \\
Mixed & 0.4--0.7  & {} & 7 & 1.4 & 482 & 2.2 $\pm$ 0.6 \\
Mixed & 0.7--0.9  & {} & 4 & 0.6 & 261 & 2.3 $\pm$ 0.8 \\
Mixed & 0.9--1.4  & {} & 1 & 0.3 & 322 & 1.2 $\pm$ 0.4 \\
N/A & 0.4--0.7  & {} & 4 & 1.5 & 482 & 1.5 $\pm$ 0.5 \\
N/A & 0.7--0.9  & {} & 1 & 1.1 & 261 & 1.0 $\pm$ 0.6 \\
N/A & 0.9--1.4  & {} & 2 & 0.9 & 322 & 1.3 $\pm$ 0.5 \\

\enddata
\label{table:dry_wet}
\tablecomments{\fpair\ has been increased by 1\% to account for losses due to catastrophic \photoz\ errors.}
\end{deluxetable}

\subsection{Comparison to Previous Work}\label{previous_work}

In this subsection, we discuss our results in the context of previous
work.  A number of close pair studies have been carried out using local
surveys with $z \lesssim 0.3$ \citep[e.g.,][]{xu04, de-propris05,
  masjedi06, patton08}.  These tend to find very low pair fractions,
often only a few percent.  Using our full sample with $\log
M_*/M_{\odot} > 10$, we find $f_{\rm pair} = 0.03 \pm 0.02$ at $z \sim
0.5$.  This is slightly above but consistent with the value of $f_{\rm
  pair} \simeq 0.01$ reported by \citet{patton08} who analyzed a
volume-limited $r$-selected SDSS sample (note that $f_{\rm pair} \approx 0.5
N_c$).  They selected pairs with $r$-band luminosity ratios greater than
$1/2$ which is more stringent than our $\approx 1/4$ $M_*$ ratio
threshold and should lead to lower \fpair\ since not all major mergers
are selected.  \citet{patton08} find no luminosity dependence in their pair
fraction estimate (with $r_{\rm sep} = $5--20 $h^{-1}$ kpc) in contrast
to the mass dependence found here.

\citet{xu04} use the combination of the 2DFRGS and 2MASS surveys to
measure both the fraction and number density of close pairs drawn from a
$K$-selected sample at $\langle z \rangle = 0.03$.  With a magnitude
threshold of $\Delta K < 1$, their study is an excellent low-$z$ analog
to ours, and they estimate $f_{\rm pair} \approx $0.02--0.03 (their pair
fraction is also defined as $2f_{\rm pair}$) in good agreement with our
lowest redshift bin.  As mentioned previously, \citet{xu04} also find
weak evidence for a higher value of \fpair\ for more massive galaxies.
Our analysis confirms this finding and demonstrates that it continues to
$z \sim 1$.

Studies at higher redshifts tend to focus on the evolution of the pair
fraction with redshift as parametrized by $m$.  This number is poorly constrained by
recent work, with $m$ varying from 0 to 4 \citep[see][]{kartaltepe07}.
Leaving evolution aside, better agreement in the average merger rate is
being achieved, however.  For example, our new results are consistent
with the initial near-IR study undertaken by \citet{bundy04} and the
work of \citet{rawat08} who perform a similar analysis using only
spectroscopic redshifts in GOODS-S.

A number of optical studies of the pair fraction also deliver similar
results, although some caution is needed in making comparisons.  In
addition to the bias in optical samples due to SF, our present work reveals a
second effect.  Depending on the selection method, optical samples may
trace different stellar mass ranges as a function of redshift.  Since we
have shown that the pair fraction depends on $M_*$, this can bias the
pair fraction at different redshifts.  With this in mind, we note that
the value of $f_{\rm pair} = $0.02--0.03 measured by both \citet{bell06a}
using COMBO-17 as well as the kinematic pair study of \citet{lin04} and
\citet{lin08} agrees with our findings here.  Similar values were also
found by \citet{kartaltepe07} and \citet{hsieh08} and, for $z \lesssim
0.8$, by \citet{de-ravel08}.

Not all merger studies find such low values, however.  Our pair fraction
is lower by a factor of 3--4 and shows less evolution compared to
\citet{le-fevre00}, likely as a result of the SF bias.  At the same
time, morphological derivations of the merger fraction tend to find
values of $f_{\rm pair} \approx 0.1$ at $0.4 < z < 1.4$, 2--3 times
higher than the results of our pair analysis
\citep[e.g.,][]{conselice03, lotz08}.  This may reflect the influence of
non-uniform star formation, minor mergers, and even flybys on galaxy
morphology as discussed with respect to Figure \ref{fig:pair_frac_morph}
\citep[also see][]{lotz08a}.

\section{Discussion: The Galaxy Merger Rate}\label{discussion}

\subsection{Insight from Galaxy Clustering}\label{clustering}

Our goal in measuring the pair fraction is to constrain the rate at
which galaxies merge, but doing so requires knowing the fraction, \cmg,
of close pairs that will eventually merge and on what timescale, \tmg,
they coalesce.  Both quantities are uncertain, but previous work has
typically assumed that roughly half of close pairs will merge with a
typical timescale of 0.5 Gyr \citep[e.g.,][]{patton00}.  Under
these assumptions, the pair fractions presented here would indicate that
mergers occur more frequently among massive, red galaxies as compared to
their less massive and bluer counterparts.  But if either \cmg\  or \tmg\ depend on mass or galaxy type, this conclusion
could be incorrect.  As we now show, evidence from galaxy clustering may
imply such a dependence.

The galaxy 2-point correlation function is typically measured on scales
larger than 100 $h^{-1}$ kpc and is usually fit as a power law $\xi =
(r/r_0)^{-\gamma}$.  By extrapolating this fit to the radii of interest
for the close pairs in this paper, one can estimate the pair fraction
predicted by clustering measurements made on larger scales.  Assuming
$\xi \gg 1$ and using our definition of \fpair,

\begin{equation}
f_{\rm pair} \approx 4 \pi n_g \int_{r_{\rm min}(1+z)}^{r_{\rm max}(1+z)} r^2 \xi(r) dr,
\end{equation}
\begin{equation}
= \frac{4 \pi n_g}{3-\gamma} r_0^\gamma (1+z)^{3-\gamma} (r_{\rm max}^{3-\gamma} - r_{\rm min}^{3-\gamma}),
\end{equation}\label{eqn:cluster}

\noindent where $n_g$ is the comoving galaxy number density of potential
companions, and the $(1+z)$ factor accounts for the fact that $r_{\rm
  min}$ and $r_{\rm max}$ are defined in physical and not comoving
coordinates.

We can use estimates of $r_0$ and $\gamma$ from various clustering
analyses to see how well the clustering signal extrapolated to small
scales agrees with the pair fractions we measure.  For example,
\citet{zehavi05} measure luminosity dependent clustering in the SDSS.
Taking their three brightest $M_r$ luminosity bins to roughly correspond to
our three stellar mass bins, we use $r_0 = 5.52, 6.16, 10.0$ in units of
$h^{-1}$ Mpc with $\gamma = 1.78, 1.85, 2.04$.  We approximate $n_g$ by
integrating the mass functions of \citet{bundy06} and use $n_g = 4, 4,
0.8$ in units of 10$^{-3}$ Mpc$^{-3}$ with $h=0.7$.  With these values
and setting $z = 0.5$ (note that SDSS has an average $z \approx 0.1$),
we find a predicted \fpair\ of 0.03, 0.06, and 0.10 for the three
``mass'' bins.  These are overplotted in the first redshift bin of the
top panel in
Figure \ref{fig:frac_mass}.  

Extending this comparison to higher redshifts is difficult because the
smaller sizes of high-$z$ samples do not adequately probe the most massive or
luminous galaxies.  Still, we can compare to the $z \sim 1$ results of
\citet{coil06} where we take $r_0 = 3.78 h^{-1}$ Mpc and $\gamma = 1.68$ from their
median $M_B \approx -20.3$ sample 1 to correspond to our lowest mass bin ($\log
M_*/M_{\odot} = $10--10.5) and $r_0 = 4.21 h^{-1}$ Mpc with $\gamma = 1.9$ from
their $M_B \approx -21$ sample 4 for our second mass bin ($\log
M_*/M_{\odot} = $10.5--11).  We find $f_{\rm pair} \approx $0.01 and
0.04 and these results are plotted in the middle redshift bin of the top
panel of Figure
\ref{fig:frac_mass}.  It should be emphasized that $M_B$ is not a
good tracer of $M_*$ and that the sample used by \citet{coil06} is not
mass-limited \citep[see][]{coil08}.  

While the comparison between \fpair\ and predictions from the 2-point
correlation function is clearly approximate, the resulting agreement is
striking.  Indeed, the lower correlation lengths of blue versus red
galaxies \citep[e.g.,][]{coil08} also seem to be consistent with the lower
pair fraction of blue hosts shown in Figure \ref{fig:pair_frac_col}.

\begin{deluxetable*}{lcccccccccc}
\tablecaption{Fractional Merger Rates, \rmg}
\tabletypesize{\footnotesize}
\tablewidth{0pt}
\tablecolumns{11}
\tablehead{
\multicolumn{4}{c}{} & \multicolumn{3}{c}{\citet{patton08}} & \colhead{} & \multicolumn{3}{c}{\citet{kitzbichler08}} \\
\cline{5-7} \cline{9-11} \\
\colhead{$z$} & \colhead{$\log{M_*}$} & \colhead{$f_{\rm pair}$} & \colhead{} & \colhead{\cmg} & \colhead{\tmg} & \colhead{\rmg} & \colhead{} &
\colhead{\cmg} & \colhead{\tmg} & \colhead{\rmg} \\
\colhead{} & \colhead{($\log{h_{70}^2 M_{\odot}}$)} & \colhead{(Method I)} & \colhead{} & \colhead{} & \colhead{(Gyr)} & \colhead{(Gyr$^{-1}$)} & \colhead{} &
\colhead{} & \colhead{($h_{70}^{-1}$ Gyr)} & \colhead{($h_{70}$ Gyr$^{-1}$)} \\
}

\startdata

\vspace*{0.005cm} \\
$0.4< z <0.7$ & $10.0< \log M_*$ & 0.03 & {} & 0.58 & 0.5 & 0.036 & {} & 1.0 & 2.0 & 0.016 \\
$0.7< z <0.9$ & $10.0< \log M_*$ & 0.05 & {} & 0.58 & 0.5 & 0.060 & {} & 1.0 & 2.0 & 0.026 \\
$0.9< z <1.4$ & $10.0< \log M_*$ & 0.06 & {} & 0.58 & 0.5 & 0.068 & {} & 1.0 & 2.0 & 0.029 \\

\vspace*{0.005cm} \\
$0.4< z <0.7$ & $10.0< \log M_* <10.5$ & 0.03 & {} & 0.55 & 0.5 & 0.034 & {} & 1.0 & 1.9 & 0.016 \\
$0.7< z <0.9$ & $10.0< \log M_* <10.5$ & 0.05 & {} & 0.55 & 0.5 & 0.057 & {} & 1.0 & 1.9 & 0.027 \\
$0.9< z <1.4$ & $10.0< \log M_* <10.5$ & 0.03 & {} & 0.55 & 0.5 & 0.038 & {} & 1.0 & 1.9 & 0.018 \\

\vspace*{0.005cm} \\
$0.4< z <0.7$ & $10.5< \log M_* <11.0$ & 0.03 & {} & 0.58 & 0.5 & 0.031 & {} & 1.0 & 1.4 & 0.019 \\
$0.7< z <0.9$ & $10.5< \log M_* <11.0$ & 0.04 & {} & 0.58 & 0.5 & 0.048 & {} & 1.0 & 1.4 & 0.030 \\
$0.9< z <1.4$ & $10.5< \log M_* <11.0$ & 0.07 & {} & 0.58 & 0.5 & 0.080 & {} & 1.0 & 1.4 & 0.050 \\

\vspace*{0.005cm} \\
$0.4< z <0.7$ & $11.0< \log M_*$ & 0.05 & {} & 0.46 & 0.5 & 0.046 & {} & 1.0 & 1.0 & 0.051 \\
$0.7< z <0.9$ & $11.0< \log M_*$ & 0.07 & {} & 0.46 & 0.5 & 0.067 & {} & 1.0 & 1.0 & 0.075 \\
$0.9< z <1.4$ & $11.0< \log M_*$ & 0.09 & {} & 0.46 & 0.5 & 0.082 & {} & 1.0 & 1.0 & 0.092 \\

\enddata
\label{table:merger_frac}
\end{deluxetable*}

\begin{deluxetable*}{lcccccccccccc}
\tablecaption{Volumetric Merger Rate Mass Function}
\tabletypesize{\footnotesize}
\tablewidth{0pt}
\tablecolumns{13}
\tablehead{
\multicolumn{2}{c}{} & \multicolumn{3}{c}{$0.4 < z < 0.7$} & \colhead{} & \multicolumn{3}{c}{$0.7 < z < 0.9$} & \colhead{} & \multicolumn{3}{c}{$0.9 < z < 1.4$} \\
\cline{3-5} \cline{7-9} \cline{11-13} \\
\colhead{Stellar Mass Range} & \colhead{} & 
\colhead{$\log{\Psi}$} & \colhead{$\log{\Psi_{\rm wet}}$} & \colhead{$\log{\sigma}$} & \colhead{} & 
\colhead{$\log{\Psi}$} & \colhead{$\log{\Psi_{\rm wet}}$} & \colhead{$\log{\sigma}$} & \colhead{} & 
\colhead{$\log{\Psi}$} & \colhead{$\log{\Psi_{\rm wet}}$} & \colhead{$\log{\sigma}$} \\
}

\startdata

$10.0< \log M_* <10.5$ & {} & -3.77 & -3.96 & 0.31 & {} & -3.69 & -3.94 & 0.25 & {} & -4.16 & -4.37 & 0.27 \\ 
$10.5< \log M_* <11$ & {} & -3.90 & -4.09 & 0.29 & {} & -3.89 & -4.14 & 0.26 & {} & -3.81 & -4.03 & 0.13 \\ 
$\log M_* > 11$ & {} & -4.51 & -4.70 & 0.28 & {} & -4.38 & -4.63 & 0.24 & {} & -4.50 & -4.72 & 0.17 \\ 

\cutinhead{Volumetric Formation Rate of Spheroidals}
$10.25< \log M_* <10.55$ & {} & -3.18 & {...} & 0.14 & {} & -3.12 & {...} & 0.09 & {} & -4.50 & {...} & 0.42 \\ 
$10.55< \log M_* <10.85$ & {} & -3.92 & {...} & 0.55 & {} & -2.91 & {...} & 0.08 & {} & -3.37 & {...} & 0.09 \\ 
$10.85< \log M_* <11.15$ & {} & -3.57 & {...} & 0.33 & {} & -3.02 & {...} & 0.10 & {} & -3.55 & {...} & 0.16 \\ 
$\log M_* > 11.15$       & {} & -4.06 & {...} & 0.52 & {} & -3.44 & {...} & 0.17 & {} & -3.60 & {...} & 0.13 \\

\enddata
\label{table:vol_merge}
\tablecomments{The values in this table are plotted in Figure \ref{fig:vol_merge}.  The symbol $\Psi$ denotes a
  rate per unit volume (per logarithmic interval) with units of $h_{70}^{-3}{\rm Mpc}^{-3} {\rm
    dex}^{-1} {\rm Gyr}^{-1}$.  It should be distinguished from $\Phi$, often used to represent mass
  or luminosity functions.  $\Psi_{\rm wet}$ indicates the approximate volumetric merger rate after
  removing dry mergers.  We use $\log{\sigma}$ to designate the associated statistical uncertainty (in dex).  Stellar
  masses have units of $h_{70}^{-2} M_{\odot}$.}
\end{deluxetable*}

\begin{figure*}
\plotone{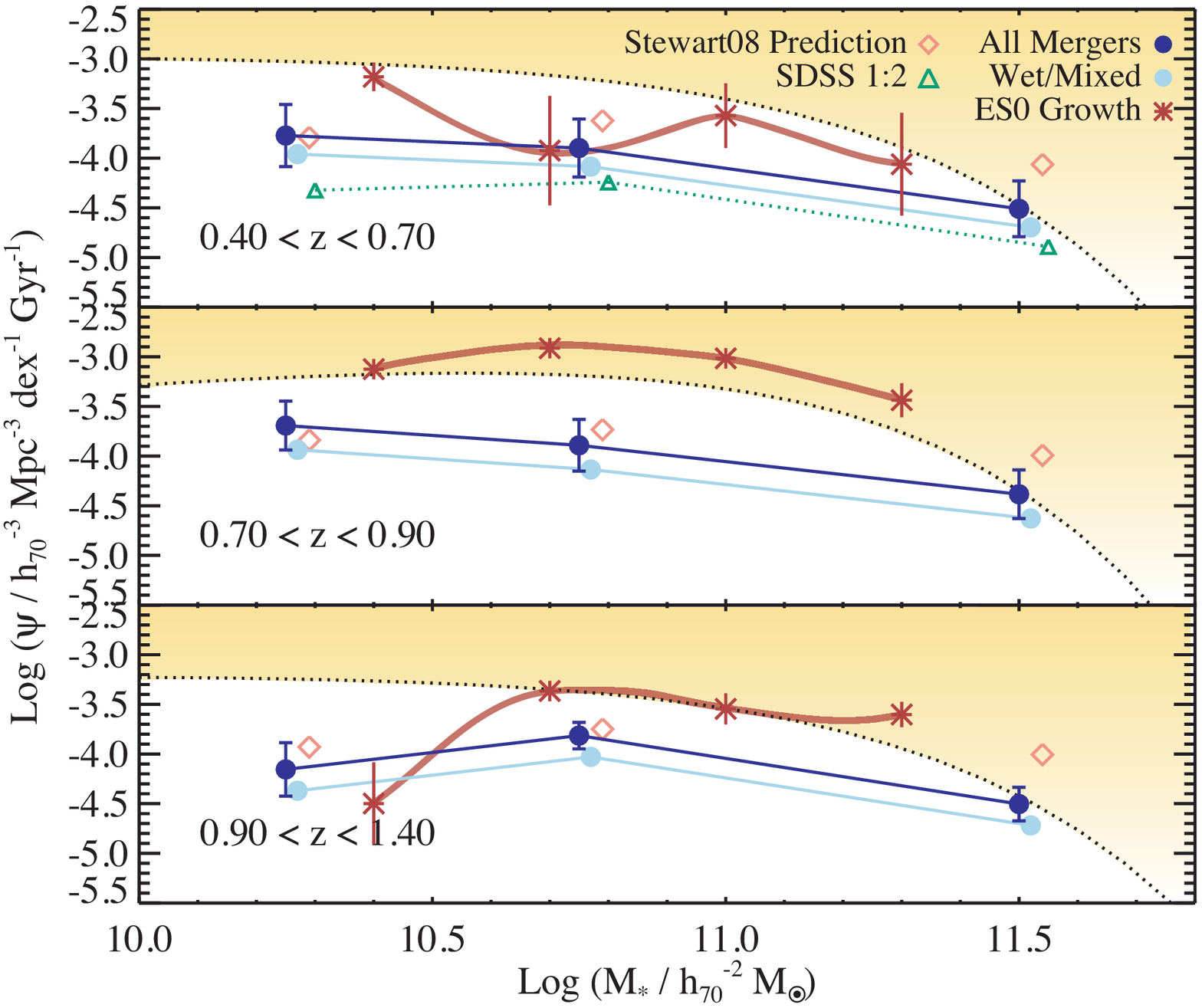}
\caption{The major merger rate mass function in three redshift bins.  The
  y-axis denotes $\Psi$, defined as a rate (per Gyr) per unit
  volume per logarithmic stellar mass interval.  Solid circles indicate
  the observed merger rate for all galaxies
determined with method I (see Fig \ref{fig:frac_mass}) and assuming the
  values of \tmg\ and \cmg\ from \citet{patton08}.  Light blue circles show
  the result of excluding the approximate fraction of dry E/S0-E/S0
  mergers, as measured in \S\ \ref{dry_wet_pairs}.  The green triangles
  in the first redshift bin reflect the luminosity-dependent volumetric
  merger rate at $z \sim 0.1$ computed for 1:2 or greater mass ratio mergers from the
  SDSS \citep{patton08}.  Predictions for galaxy merger rates based on
  cosmological simulations from \citet{stewart08} are plotted as open
  diamonds.  Red asterisk symbols show the rate of growth
  in the number of E/S0 galaxies based on the sample from
  \citet{bundy05}.  The rate of merging appears to be too low to fully
  account for the generation of new early-type galaxies.  The dotted
  line in each panel indicates the average merger rate
  obtained if every galaxy experiences one merger over the range, $0.4 <
  z < 1.4$.  Rates lying in the shaded region above these lines have a strong impact
  on the population, while those below have a minimal effect. 
  \label{fig:vol_merge}}
\end{figure*}

This suggests that the connection between \fpair\ and clustering is a
strong one and that the trends observed here may be expected given the
way the correlation function increases with galaxy luminosity, stellar
mass, and for red, early-type systems \citep[e.g.,][]{norberg02,
  zehavi05, li06, meneux06, coil06, coil08, meneux08}.  In other words,
the trends we observe with \fpair\ may be driven by regions of higher
density which tend to host more massive, red-sequence systems.

\subsection{Estimates of the Merger Timescale from Simulations}

Perhaps the best way to estimate \cmg\ and \tmg\ is through detailed
simulations.  Unfortunately, fully numerical merger simulations are
expensive and can only be performed for a handful of systems
\citep[e.g.,][]{boylan-kolchin07}.  Cosmological simulations, meanwhile,
are limited in resolution and require various analytic assumptions
regarding dynamical friction, tidal interactions, and other details.
Both \citet{kitzbichler08}
and \citet{patton08} use the Millennium Simulation to determine the
frequency and timescales of merging pairs, and after accounting for the
fact that \citet{kitzbichler08} absorb \cmg\ into their estimate of
$\langle T_{\rm merge} \rangle$ \citep[see][]{patton08}, the two studies derive
similar results on average.

\citet{kitzbichler08} find that $\langle T_{\rm merge} \rangle$ depends
inversely on stellar mass as $M_*^{-0.3}$.  This actually {\em enhances}
the mass dependence of the merger rate beyond what we have observed for
\fpair.  In other words, the higher \fpair\ for massive galaxies
translates into an even faster merger rate.  \citet{patton08}, on the
other hand, assume that \tmg, defined for real-space pairs destined to
merge, is a roughly constant 0.5 Gyr and instead investigate the
dependence of \cmg\ (defined as $f_{\rm 3D}$ in their paper) on
luminosity.  As might be expected from the \citet{kitzbichler08}
analysis, they find that \cmg\ increases with luminosity, resulting in
more efficient merging for more luminous pairs (equivalent to a lower
value of $\langle T_{\rm merge} \rangle$ for more massive galaxies).
However, it is interesting that their most luminous pairs ($M_r \approx
-22.8$ or $\log{M_*/M_{\odot}} \approx 11.2$) show a drop in \cmg\ which
may be evidence of the clustering effect discussed above.  The
\citet{patton08} values therefore have a mild effect and translate the
trend in \fpair\ into a slightly more moderate mass dependence for the
merger rate. \\

\subsection{The Merger Rate and Merger Rate Mass Function}

Keeping in mind the substantial uncertainty involved with converting
from \fpair\ to a merger rate, we derive merger rates using the
assumptions in both \citet{kitzbichler08} and \citet{patton08}.  In
Table \ref{table:merger_frac} we present the fractional galaxy merger
rate, \rmg, defined as ${\cal R}_{\rm mg} \equiv C_{\rm mg} f_{\rm
  pair}/T_{\rm mg}$.  \rmg\ can be thought of as the fraction of mergers
per unit time within a sample of galaxies.  We take the three most
luminous bins in \citet{patton08} to correspond to our three mass bins
and use the fitting formula for \tmg\ (with \cmg\ set to 1.0) from
\citet{kitzbichler08}.

With \rmg\ determined, we use our mass-limited sample to
self-consistently compute the volumetric merger rate (the number of
mergers per unit time per unit comoving volume) as a function of the
stellar mass of the primary or host galaxy.  We call this the {\it
  merger rate mass function} and denote it using the variable, $\Psi$.  The merger rate MF can be directly
compared to differential evolution in the MFs of galaxies of various
type, making it a valuable tool for evaluating the role of mergers in
mass assembly and galaxy evolution.

The merger rate mass function is determined by multiplying the MF of the
different host galaxy samples drawn from our survey (see
\ref{method_MF}) by the relevant value of \rmg.  In what follows, we use
the values of \rmg\ determined from the merger timescales and
efficiencies reported by \citet{patton08} and listed in Table
\ref{table:merger_frac}.  Using the values from \citet{kitzbichler08}
leads to even higher merger rates (by $\sim$10\%) for the most massive galaxies.  Our results
are plotted as the dark solid circles in Figure
\ref{fig:vol_merge}.  The error bars reflect the uncertainty in \fpair\
which dominates over the statistical uncertainty in the computed number
densities.  Both \fpair\ and the corresponding MFs have been determined
using the same GOODS N+S dataset described in \S\ \ref{data}.  It should
be emphasized, however, that both suffer from cosmic variance, although
this problem is mitigated by combining two independent fields.  We
estimate that cosmic variance affects the data plotted in Figure
\ref{fig:vol_merge} at the 0.2 dex level.  The largest single source of
uncertainty lies in the assumed values of \tmg\ and \cmg\, however.

The volumetric merger rate of the SDSS sample ($z \sim 0.1$) measured in
different luminosity bins (with pair separation 5--20 $h^{-1}$ kpc) by
\citet{patton08} is roughly plotted as a function of mass in the low-$z$
bin of Figure \ref{fig:vol_merge}.  Given the lack of strong redshift
evolution in our sample, the lower merger rate from SDSS may simply
reflect the fact that \citet{patton08} only consider merging pairs with
an optical luminosity ratio greater than $1/2$.  Additional differences
may arise in how \cmg\ and \tmg\ are applied to our different pair
selection methods.  Also shown in Figure \ref{fig:vol_merge} are
predictions for the 1:4 or greater $M_*$ ratio galaxy merger rate based
on the cosmological N-body simulations and halo modeling performed in
\citet{stewart08}.  These are multiplied by the observed MFs to derive
the volumetric rates plotted.  The agreement is remarkable, especially
given the uncertainties in \tmg.  There is a hint that the merger rate
derived from observations falls below the model predictions, especially
at the highest masses.

\subsection{Hierarchical Mass Assembly Through Merging}\label{mass_assembly}

Figure \ref{fig:vol_merge} provides key insight about the nature of
galaxy mass assembly.  The dotted line in the figure provides a
benchmark useful for gauging the impact of mergers on the galaxy
population at different masses.  It is determined by dividing the galaxy
abundance at each redshift \citep[from the MFs of][]{bundy06} by 4.5
Gyr, the amount of time spanning our redshift range, $0.4 < z < 1.4$.
It is therefore the average event rate that would be obtained if every
galaxy experienced exactly one merger over our redshift range.  Rates
that fall above this benchmark (in the shaded region of the plot)
indicate a strong impact on the population.  Rates that fall below
correspond to processes with a minimal impact.

Figure \ref{fig:vol_merge} shows that for galaxies with $M_* \lesssim 10^{11} M_{\odot}$, the
observed merger rate lies below our benchmark, indicating that
major mergers have {\em not} contributed significantly to the assembly
history of such galaxies since $z \sim 1$.  For more massive galaxies
the situation changes.  In the highest mass bin, the merger rate is roughly on par with
the benchmark rate, demonstrating that major mergers play a larger
role in the recent assembly of such galaxies.

We can quantify this trend with the ``merger remnant fraction,'' that
is the fraction of systems that have undergone major mergers in the
different mass bins.  This number is determined by integrating \rmg\
over $0.4 < z < 1.4$.  Using the results from method I, we find that at
the highest masses, $\sim$30\% of galaxies experience a major
merger.  At the lowest masses probed, this number drops to 10--15\%.
These estimates are approximate because we have not considered transfers
across mass bins as a result of mergers and SF
\citep[see][]{drory08} and because of the large uncertainties in \cmg\
and \tmg\ \citep[also see][]{bell06}.  Still, they reinforce the
increasing importance of mergers on the assembly history of higher mass
galaxies at $z \sim 1$.  

Similarly, we can compare the volumetric merger rate to the average SFR
measured as a function of $M_*$ \citep[e.g.,][]{drory08}.  While
measurements of SFR$(M_*)$ remain uncertain, this exercise indicates
that the mass growth since $z \lesssim 1.5$ of galaxies with $\log
M_*/M_{\odot} > 11$ is almost completely dominated by merging.  For
systems with $\log M_*/M_{\odot} \sim 10$, however, new growth from SF
amounts to roughly 10 times the stellar mass accreted through major
mergers, clearly indicating that SF is a far more important source of
growth at lower masses.  Considering the full mass range, $\log
M_*/M_{\odot} \ge 10$, the mass accreted in major mergers over $0.4 < z
< 1.4$ amounts to $\sim$15\% of the total stellar mass density of
systems with $\log M_*/M_{\odot} \ge 10$ at $z \sim 0.8$.  For $\log
M_*/M_{\odot} \ge 11$, mass accretion through major mergers over the
same redshift range accounts for $\sim$25\% of $\rho_*$ at $z \sim 0.8$.

Qualitatively, our observations of an increase in the merger fraction
with mass echo the hierarchical assembly of dark matter halos as
determined with $\Lambda$CDM N-body simulations.  In an analysis of the
Millennium Simulation, \citet{fakhouri08}, for example, find that the
{\em halo} merger rate (for halo mass ratios greater than 1:3) rises by
$\sim$30\% from $M_{\rm DM} = 10^{12}$ to $10^{13}$.  This mass
dependence is significantly weaker but in the same sense as the factor
of $\sim$2 increase with mass we observe in the galaxy merger fraction
among the $10 < \log{M_*/M_{\odot}} < 11.5$ galaxies thought to populate
such halos.

The enhanced mass dependence in the galaxy merger rate may be due to the
way halos are occupied (details forthcoming in Hopkins et al., in prep), specifically the
inferred peak in the $M_*/M_{\rm DM}$ ratio at $M_{\rm DM} \sim 10^{12}
M_{\odot}$ \citep[e.g..][]{wang06, mandelbaum06} which roughly
corresponds to the low-mass end of the range probed by our sample.  In
this scenario, the major merger rate of halos and galaxies should be
directly related near the $M_*/M_{\rm DM}$ peak.  But at higher masses
$M_*$ declines with respect to halo mass.  This means that minor halo
mergers can contribute to {\em major} galaxy mergers since less massive
halos, being closer to the $M_*/M_{\rm DM}$ peak, have larger $M_*$
fractions.  Because there are always more minor than major halo mergers,
this enhances the observed fraction of major galaxy mergers at higher
masses.  If our observations could probe below the peak, this
interpretation would predict a suppressed galaxy major merger rate
relative to the halo major merger rate since here some fraction of major
halo mergers would host galaxies with a minor mass ratio.

  Indeed, this qualitative picture is borne out in detail by
  \citet{stewart08}.  As discussed above, the galaxy major merger rate
  predicted from their model based on N-body simulations is in
  remarkable agreement with our observations (see Figure
  \ref{fig:vol_merge}).

\subsection{Are Spheroidals Formed in Major Mergers?}\label{vol_merge}

Another important insight gained from Figure \ref{fig:vol_merge} is the
role of mergers in the formation of elliptical and red sequence
galaxies.  The red asterisks in this plot show the formation rate (the
rate at which the abundance increases) of visually classified spheroidal
systems, as determined from the sample of \citet{bundy05}.  Error bars
are Poissonian and do not include cosmic variance which we estimate
enters at the $\sim$0.2 dex level \citep[see][]{stringer08}.

Figure \ref{fig:vol_merge} shows that the formation rate of new
spheroidal galaxies is greater than the merger rate at nearly all masses
and redshifts probed.  The disagreement increases when only wet or mixed
mergers are considered---only these can create new spheroidals.  The
light blue points show the merger rate mass function after the
approximate fraction of dry spheroidal-spheroidal mergers (as measured
in \S\ \ref{dry_wet_pairs}) is subtracted.  This result was anticipated
by an analysis of the expected number of major halo mergers as measured
in the Millennium Simulation which is also unable to account for the
rate at which new spheroidals appear \citep{bundy07, genel08}.  A
similar conclusion applies to the inability of major mergers to account
for the formation of red sequence galaxies which increase at a similar
rate as the spheroidal population \citep[e.g.,][]{bell07}.

The discrepancy between major (wet) mergers and the formation of
spheroidals shown in Figure \ref{fig:vol_merge} ranges from factors of
$\sim 3$--12.  Simply interpreted, this requires mechanisms in addition
to merging to make spheroids.  How strong is this requirement?  One
concern is cosmic variance which could affect both the merger and
spheroidal formation rates---we estimate the discrepancy could be
overestimated by no more than a factor of $\sim$2 as a result
\citep{stringer08}.  At the same time, the gap could be closed to some
extent if the merger timescale were less than $\approx$0.5 Gyr
\citep[e.g.,][]{lotz08a}.  Indeed, a comparison of various theoretical
treatments concludes that current predictions for \tmg\ are again
uncertain at the factor of $\sim$2 level (Hopkins et al., in prep).
Furthermore, our spheroidal classification, while dominated by
ellipticals, includes S0's and some Sa galaxies\footnote{The fact that
  it is so difficult to discriminate between such galaxies at $z \gtrsim
  0.4$ is the reason we group them into a single ``spheroidal'' category
  \citep{brinchmann98}.}.  The large bulges in such galaxies could be
built in mergers with mass ratios less than 1:4, increasing the number
of relevant mergers by $\sim30$\%.  Finally, multiple minor mergers can
also drive morphological evolution.  We would expect a factor of $\sim$2
times more 1:10 minor mergers than the number of observed 1:4 mergers
\citep[e.g.,][]{stewart08}.  If $\sim$3 such minor mergers have the
equivalent morphological impact as one 1:4 merger \citep{bournaud07,
  hopkins08}, minor mergers can account for an additional $2/3 \approx
70$\% of the morphological transformations needed to form spheroidals.

Assuming these potential systematic effects conspire in the same way and
including minor mergers, it would be possible to reduce the discrepancy
by a factor of $\sim$6.  This would lend stronger support to the merger
hypothesis and motivate a more precise treatment, but would still leave
some tension at $z \sim 0.8$.  For this reason and given the current
observations and most recent theoretical assumptions, we conclude that
other mechanisms in addition to merging may be needed to explain the
transformation of disk-like galaxies into spheroidals
\citep[see][]{parry08} and the quenching of star formation.

\section{Summary}\label{summary}

Using a deep \K-selected catalog comprised of data from the GOODS
fields, we have presented an analysis of the close pair fraction and
implied merger rate of galaxies drawn from a mass-limited sample over
the redshift range $0.4 < z < 1.4$.  Using two methods of estimation, we
find a relatively low pair fraction of $\sim$4\% with a redshift
dependence of $(1+z)^{1.6 \pm 1.6}$.  Our analysis strongly supports the
basic conclusion of a low merger rate since $z \sim 1$, as deduced by
several other recent surveys.

Our key observational finding is that the pair fraction of galaxies with
$M_* > 10^{11}$\msun\ is higher ($f_{\rm pair} \sim 5$--9\%) than the
corresponding pair fraction of lower mass systems ($f_{\rm pair} \sim
2$--4\% for $M_* \sim 10^{10}$\msun).  In addition, red systems and
galaxies with spheroidal morphologies appear more likely to have
companions than their blue or disk-like counterparts.  This is in line
with extrapolations of the correlation function to small scales.  We
find that the fraction of host galaxies classified as irregular (often
more than half of hosts with $M_* < 3 \times 10^{10}$) is larger than
the fraction of irregulars in the parent population (typically
10--20\%).  This supports the notion that orbiting companions can
distort the primary galaxy even before a merger has occurred.

Dry mergers are only significant at the highest masses in our sample,
$M_* > 10^{11}$\msun, but this appears to result from the fact that most
galaxies in this mass range have already become red, early-type systems
by $z \sim 1$.  At lower masses ($M_* \sim 10^{10}$\msun) the fraction
of red galaxies that host companions is consistent with zero, and we
thus conclude that dry mergers alone are incapable of fully explaining
the build-up of the red sequence, especially at $M_* \approx 2$--$3
\times 10^{10}$\msun\ where the formation of red galaxies is
particularly rapid.  In support of this conclusion, we also find that
the number of dry pairs accounts for $\sim$50\% of all pairs in our mass
range and evolves in way that is consistent with pairs drawn randomly
from the parent population.  The emerging picture is one in which dry
mergers are not the main driver for red sequence growth but instead
become increasingly common as a result of it.

By adopting estimates for the close pair merger efficiency and
timescale, we determine the mass-dependent volumetric merger rate.  From
a comparison to the formation rate of galaxies with spheroidal
morphology at the same redshifts, we conclude that the major merger rate
is too low to fully explain the formation of spheroidal (and red sequence)
galaxies since $z \sim 1$.

In terms of mass assembly, major mergers have a strong impact on
galaxies with $M_* > 10^{11}$\msun, $\sim$30\% of which have experienced
a major merger during the time spanning our redshift range.  For less
massive galaxies ($M_* \lesssim 3 \times 10^{10}$\msun), such mergers
play a less significant role, affecting less than 10\% of the
population.  For systems at such intermediate and lower masses, star
formation is likely to be of much greater importance in driving growth.
The higher merger fraction observed for the more massive galaxies in our
sample provides direct support for hierarchical assembly and agrees with
recent galaxy models based on N-body simulations, even if additional
processes are required to explain the top-down nature of star formation
downsizing and the formation of spheroidal galaxies.


\vspace{2cm}
\section{Acknowledgments}

We would like to thank Ichi Tanaka for his help in obtaining our Subaru
observations and Yen-Ting Lin and Phil Hopkins for useful discussions.
MF is supported by a Grant-in-Aid of the Ministry of Education of Japan.
RSE acknowledges the financial support of the Royal Society.

\appendix
\section{MOIRCS Reductions}\label{app:moircs}

We provide additional information on the MOIRCS data reduction, which
was carried out using a modified version of the MCSRED package written
by Ichi Tanaka.  Simply using the package as written often led to the
appearance of fringes across the entire reduced frame, especially for
detector 2.  More than half of our pointings were strongly affected, and
the typical peak-to-trough fringe height was a few percent of the
background level.

We traced the cause to two problems.  First, strong fringing in one or
two frames in an image set would be imprinted into other frames via the
skyflat procedure and second, skyflats in MCSRED are made by averaging
all images in a set.  The flat field can vary over timescales of
minutes, however, and changes (as well as the fringing) may be related
to dithering the telescope.  Thus a ``running'' sky flat that includes
only some images taken before and after the science frame is desirable.
We wrote a separate procedure which constructs running skyflats to
flatfield the data and allows frames with bad fringing to be removed.
The image registration portion of MCSRED also caused occasional
problems, with some images being excluded from the final mosaic.  By
default, MCSRED calls the IRAF routine {\sc xyxymatch} with the
automated ``triangles'' registration option, which can sometimes fail
for certain frames.  When this happened we identified the positions of
three stars common to all frames in an image set and supplied these to
{\sc xyxymatch} using the ``tie points'' method.

Careful inspection of the MOIRCS images revealed a curious pattern of
noise spikes (although ``noise worms'' is a better description since
they tend to be elongated) that often appear within $\sim$10\arcsec\ of
bright sources (often stars) with $K_s \lesssim 16$.  These objects are
often detected as sources with $K_s \approx 23$ and, when they appear
near galaxies, can be mistaken for fainter companions.  We identified
133 such noise worms (about 1\% of MOIRCS detections) and removed them from
the photometric catalogs.

\clearpage


\clearpage

\renewcommand{\thetable}{\arabic{table}}
\setcounter{table}{0}
\begin{landscape}
\begin{deluxetable}{lccccccccccccccccccccc}
\tablecaption{Results for \fpair}
\tabletypesize{\scriptsize}
\tablewidth{0pt}
\tablecolumns{22}
\tablehead{
\multicolumn{3}{c}{} & \multicolumn{4}{c}{$\log M_*/M_{\odot} > 10$} & \colhead{} & \multicolumn{4}{c}{$10 < \log M_*/M_{\odot} < 10.5$}& \colhead{} &
                       \multicolumn{4}{c}{$10.5 < \log M_*/M_{\odot} < 11$} & \colhead{} & \multicolumn{4}{c}{$\log M_*/M_{\odot} > 11$} \\ 
\cline{4-7} \cline{9-12} \cline{14-17} \cline{19-22} \\
\colhead{Sample} & \colhead{$z$} & \colhead{} & \colhead{$N_P$} & \colhead{$N_{corr}$} & \colhead{$N_{gal}$} & \colhead{\fpair\ (\%)} & 
                                   \colhead{} & \colhead{$N_P$} & \colhead{$N_{corr}$} & \colhead{$N_{gal}$} & \colhead{\fpair\ (\%)} & 
                                   \colhead{} & \colhead{$N_P$} & \colhead{$N_{corr}$} & \colhead{$N_{gal}$} & \colhead{\fpair\ (\%)} & 
                                   \colhead{} & \colhead{$N_P$} & \colhead{$N_{corr}$} & \colhead{$N_{gal}$} & \colhead{\fpair\ (\%)} \\
\cline{1-22} \\
\multicolumn{22}{c}{Method I}
}

\startdata
All & 0.4--0.7  & {} & 60 & 43.9 & 514 & 3 $\pm$ 2 & {} & 37 & 29.4 & 250 & 3 $\pm$ 2 & {} & 18 & 12.7 & 200 & 3 $\pm$ 2 & {} & 5 & 1.8 & 64 & 5 $\pm$ 2 \\
All & 0.7--0.9  & {} & 56 & 36.5 & 380 & 5 $\pm$ 3 & {} & 33 & 23.6 & 183 & 5 $\pm$ 3 & {} & 16 & 10.4 & 136 & 4 $\pm$ 3 & {} & 7 & 2.6 & 61 & 7 $\pm$ 3 \\
All & 0.9--1.4  & {} & 126 & 79.8 & 793 & 6 $\pm$ 2 & {} & 53 & 41.7 & 326 & 3 $\pm$ 2 & {} & 54 & 30.9 & 333 & 7 $\pm$ 2 & {} & 19 & 7.1 & 133 & 9 $\pm$ 2 \\

Blue & 0.4--0.7  & {} & 33 & 24.5 & 268 & 3 $\pm$ 3 & {} & 27 & 20.0 & 183 & 4 $\pm$ 3 & {} & 6 & 4.2 & 75 & 2 $\pm$ 3 & {} & 0 & 0.3 & 10 & -3 $\pm$ 3 \\
Blue & 0.7--0.9  & {} & 33 & 21.9 & 212 & 5 $\pm$ 3 & {} & 28 & 17.4 & 142 & 7 $\pm$ 3 & {} & 4 & 4.0 & 58 & 0 $\pm$ 3 & {} & 1 & 0.4 & 12 & 5 $\pm$ 3 \\
Blue & 0.9--1.4  & {} & 87 & 59.2 & 574 & 5 $\pm$ 2 & {} & 47 & 38.4 & 304 & 3 $\pm$ 2 & {} & 33 & 17.8 & 206 & 7 $\pm$ 2 & {} & 7 & 3.1 & 63 & 6 $\pm$ 2 \\

Red & 0.4--0.7  & {} & 27 & 19.4 & 246 & 3 $\pm$ 3 & {} & 10 & 9.4 & 67 & 1 $\pm$ 3 & {} & 12 & 8.5 & 125 & 3 $\pm$ 3 & {} & 5 & 1.5 & 54 & 6 $\pm$ 3 \\
Red & 0.7--0.9  & {} & 23 & 14.6 & 168 & 5 $\pm$ 4 & {} & 5 & 6.2 & 41 & -3 $\pm$ 4 & {} & 12 & 6.3 & 78 & 7 $\pm$ 4 & {} & 6 & 2.2 & 49 & 8 $\pm$ 4 \\
Red & 0.9--1.4  & {} & 39 & 20.5 & 219 & 8 $\pm$ 4 & {} & 6 & 3.3 & 22 & 12 $\pm$ 4 & {} & 21 & 13.1 & 127 & 6 $\pm$ 4 & {} & 12 & 4.1 & 70 & 11 $\pm$ 4 \\

\cutinhead{Method II}
All & 0.4--0.7  & {} & 21 & 8.2 & 514 & 3 $\pm$ 1 & {} & 8 & 5.5 & 250 & 2 $\pm$ 1 & {} & 10 & 2.3 & 200 & 5 $\pm$ 1 & {} & 3 & 0.3 & 64 & 5 $\pm$ 1 \\
All & 0.7--0.9  & {} & 23 & 10.1 & 380 & 4 $\pm$ 2 & {} & 10 & 6.6 & 183 & 3 $\pm$ 2 & {} & 10 & 2.9 & 136 & 6 $\pm$ 2 & {} & 3 & 0.6 & 61 & 5 $\pm$ 2 \\
All & 0.9--1.4  & {} & 45 & 22.5 & 793 & 4 $\pm$ 1 & {} & 15 & 10.9 & 326 & 2 $\pm$ 1 & {} & 20 & 9.6 & 333 & 4 $\pm$ 1 & {} & 10 & 2.0 & 133 & 7 $\pm$ 1 \\

\enddata
\label{table:fpair}
\tablecomments{We define $N_P$ as the raw number of pairs, while
  $N_{corr}$ is the estimated number of contaminants.  The number of
  host galaxies is given by $N_{gal}$.   Method II \fpair\ values have been corrected upward by
  1\% to account for catastrophic \photoz\ errors.}
\end{deluxetable}

\clearpage
\end{landscape}

\end{document}